\documentclass[prl,preprint,superscriptaddress]{revtex4-1}
\usepackage[pdftex,colorlinks=true,bookmarks=false,citecolor=blue,urlcolor=blue,linkcolor=blue]{hyperref}
\usepackage{pdfpages}
\usepackage{amsmath}
\usepackage{amssymb}
\usepackage{graphicx}

\usepackage{hyperref}
\usepackage{bm}
\usepackage{color}
\usepackage{pifont}
\usepackage{graphicx}% Include figure files
\usepackage{dcolumn}% Align table columns on decimal point
\usepackage{bm}% bold math
\usepackage{natbib}
\usepackage{amsmath}
\usepackage{bbm}
\usepackage{amsthm,amsmath,amssymb,esint}
\usepackage{mathrsfs}
\usepackage{verbatim}
\usepackage{physics}
\usepackage{BOONDOX-calo}
\usepackage[normalem]{ulem}
\usepackage{textcomp}
\usepackage{float}
\newcommand\figsize{0.85}

\makeatletter
\AtBeginDocument{\let\LS@rot\@undefined}
\makeatother

\begin{document}
	%%%%%%%%%%%%%%%% Title Section Begin %%%%%%%%%%%%%%%
	\title{Electromagnetically Induced Transparency in Strongly Relativistic Regime}% Force line breaks with \\

	\affiliation{Beijing National Laboratory for Condensed Matter Physics, Institute of Physics, CAS, Beijing 100190, China}
	\affiliation{Department of Physics and Beijing Key Laboratory of Opto-electronic Functional Materials and Micro-nano Devices, Renmin University of China, Beijing 100872, China}
	\affiliation{School of Physical Sciences, University of Chinese Academy of Sciences, Beijing 100049, China}
     \affiliation{Key Laboratory of Quantum State Construction and Manipulation (Ministry of Education), Renmin University of China, Beijing, 100872, China}
	\affiliation{IFSA Collaborative Innovation Center, Shanghai Jiao Tong University, Shanghai 200240, China}
	\affiliation{Songshan Lake Materials Laboratory, Dongguan, Guangdong 523808, China}
	\affiliation{Key Laboratory for Laser Plasmas (MoE) and School of Physics and Astronomy, Shanghai Jiao Tong University, Shanghai 200240, China}

	\author{Tie-Huai Zhang}
	\affiliation{Beijing National Laboratory for Condensed Matter Physics, Institute of Physics, CAS, Beijing 100190, China}%
	\affiliation{School of Physical Sciences, University of Chinese Academy of Sciences, Beijing 100049, China}%

	\author{Wei-Min Wang}%
	\email{weiminwang1@ruc.edu.cn}
	\affiliation{Department of Physics and Beijing Key Laboratory of Opto-electronic Functional Materials and Micro-nano Devices, Renmin University of China, Beijing 100872, China}%
     \affiliation{Key Laboratory of Quantum State Construction and Manipulation (Ministry of Education), Renmin University of China, Beijing, 100872, China}
	\affiliation{IFSA Collaborative Innovation Center, Shanghai Jiao Tong University, Shanghai 200240, China}

	\author{Yu-Tong Li}
	\email{ytli@iphy.ac.cn}
	\affiliation{Beijing National Laboratory for Condensed Matter Physics, Institute of Physics, CAS, Beijing 100190, China}%
	\affiliation{School of Physical Sciences, University of Chinese Academy of Sciences, Beijing 100049, China}
	\affiliation{IFSA Collaborative Innovation Center, Shanghai Jiao Tong University, Shanghai 200240, China}
	\affiliation{Songshan Lake Materials Laboratory, Dongguan, Guangdong 523808, China}
	
	\author{Jie Zhang}
	\affiliation{Beijing National Laboratory for Condensed Matter Physics, Institute of Physics, CAS, Beijing 100190, China}
	\affiliation{IFSA Collaborative Innovation Center, Shanghai Jiao Tong University, Shanghai 200240, China}
	\affiliation{Key Laboratory for Laser Plasmas (MoE) and School of Physics and Astronomy, Shanghai Jiao Tong University, Shanghai 200240, China}

	\date{\today}% It is always \today, today,
	
	\begin{abstract}
		Stable transport of laser beams in highly over-dense plasmas is of significance in fast ignition of inertial confinement fusion, relativistic electrons generation, and powerful electromagnetic emission, but hard to realize. Early in 1996, Harris proposed an electromagnetically induced transparency (EIT) mechanism, analogous to the concept in atom physics, to transport a low-frequency (LF) laser in over-dense plasmas aided by a high-frequency pump laser. However, subsequent investigations show that EIT cannot occur in real plasmas with boundaries. Here, our particle-in-cell simulations show that EIT can occur in strongly-relativistic regime and results in stable propagation of a LF laser in bounded plasmas with tens of its critical density. A relativistic three-wave coupling model is developed, and the criteria and frequency passband for EIT occurrence are presented. The passband is sufficiently wide in strongly-relativistic regime, allowing EIT to work sustainably. Nevertheless, it is narrowed to nearly an isolated point in weakly-relativistic regime, which can explain the quenching of EIT in bounded plasmas found in previous investigations.

	\end{abstract}
	
	%\keywords{Magnetic diffusion, dense plasma}%Use showkeys class option if keyword
	\maketitle
	%%%%%%%%%%%%%%%% Title Section End %%%%%%%%%%%%%%%
	
	Transport of ultra-intense lasers in plasmas in different density ranges is crucial both in fundamental researches of laser-plasma physics and diverse applications, including particle acceleration \cite{Esarey2009,Macchi2013}, brilliant X-ray emission \cite{Armstrong2019,Schoenlein2019}, relativistic electrons generation \cite{Rosmej2019,Chopineau2019}, and inertial confinement fusion (ICF) \cite{Nuckolls1972,Kodama2001,Zhang2020}. 
	In recent laser-plasma experiments, the laser beams with normalized field strength reaches $a>1$ and even $a\gg 1$ ($a=eE/m c \omega$) have been widely adopted \cite{Miquel2019,Meyerhofer2010,Jiao2018}, where $e$ and $m$ are electron charge and mass, and $\omega$ and $c$ are laser frequency and speed in vacuum. 
	In such a laser field, the intrinsic relativistic nonlinearity would lead to the well-known relativistic transparency (RT) effect \cite{Kaw1970,Palaniyappan2012}, where $\gamma m$ ($\gamma$ is the Lorentz factor) increases, and over-dense plasmas could not shield the laser field. Since RT effect provides a feasible path for deeper penetration of laser fields and enhancing their interaction with plasmas, it has become one of the central issues on the intense lasers interacting with high-density targets \cite{Willingale2009,Weng2012,Frazer2020,Stark2015}.
	                                                                                 	
	Multiple laser beam incidence in plasmas is anticipated to give rise to some novel effects, one of which is electromagnetically induced transparency (EIT) \cite{Harris1996}. Analogous to the concept of EIT in atom physics \cite{Fleischhauer2005}, it means that a low-frequency (LF) wave (also called Stokes wave) with $\omega_1<\omega_{cut}$ can still propagate through plasmas when aided by a pump wave with a frequency $\omega_0>\omega_{cut}$, where the cut-off frequency $\omega_{cut}=\omega_p=\sqrt{4\pi e^2 n_e/m}$ for unmagnetized cold plasmas, owing to the interference between the two waves \cite{Harris1996}. The current induced by the LF wave tends to be canceled by the beat current of pump wave and plasma oscillation when $\omega_0-\omega_p<\omega_1$, which results in transparency of the LF wave \cite{Harris1996,Gordon2000}. Previous investigations on EIT in plasma physics primarily focused on weakly or non-relativistic laser cases ($a\ll 1$). Those investigations have shown that EIT has stringent prerequisites, such as a narrow passband of the LF wave frequency \cite{Matsko1998,Ersfeld2002} or a strongly-magnetized plasma background where the cyclotron frequency is comparable to the EM waves \cite{Hur2003}. By considering the three-wave coupling processes where the anti-Stokes wave $\omega_2=2\omega_0-\omega_1$ is introduced, Gordon \emph{et al.} derived the dispersion relationship of the Stokes wave under the condition $|a_1|,|a_2|\ll a_0\ll 1$ \cite{Gordon2000}. However, the transport of the Stokes wave of Stokes-dominated cases ($|a_2|\ll|a_1|$) in real plasmas with boundaries was not observed in particle-in-cell (PIC) simulations. Consequently, a negative assertion was put forth, stating that EIT would not work in bounded plasmas \cite{Gordon2000a}.
	
	Here, our PIC simulations show that EIT can work in bounded plasmas in strongly-relativistic regime, which is explained by a relativistic three-wave coupling model. The passband allowing transport of the LF wave is broadened with the increase in the pump laser intensity. In weakly-relativistic regime, the passband is too narrow and the phase matching among the pump, Stokes, and anti-Stokes waves is easy to be broken when the plasma density or $\omega_p$ is perturbed by the laser interaction. As the passband becomes sufficiently wide, the phase matching can be robust against laser-driven density perturbation, which therefore can result in stable propagation of the LF wave in over-dense plasmas in strongly-relativistic regime. Furthermore, EIT is triggered by the interference of the pump and Stokes waves, which requires the wave polarization is parallel. However, the parallel polarization is not necessary for conventional RT when the pump intensity is much higher than the Stokes one, which can be distinguished from EIT (note that anisotropic-momentum distribution could affect RT \cite{Stark2015,Arefiev2020}). This study clarifies the long-standing theoretic problem of EIT in plasma physics and can be applied in DCI \cite{Zhang2020} or fast ignition \cite{Tabak1994,Kodama2001,Wang2015b} to improve the electron spectrum and yield \cite{Zhang2023} by mixing the second-harmonic and fundamental lasers. 

	\emph{Simulation setup}.\textendash The 1D-3V particle-in-cell (PIC) code KLAPS \cite{Wang2015a} is utilized to investigate the EIT effect. Here, the pump laser wave is denoted by a footnote ``0", while the Stokes and anti-Stokes waves by ``1" and ``2", respectively. All the wave envelopes are set to be uniform after a sin-shaped rise for $5T_0$. The waves are p-polarized (along the $y$ direction), and the frequency relationship is taken as $\omega_1=0.4\omega_0$. The simulation domain is $40\lambda_0$ (2560 cells) in the $x$-direction and the waves are injected from the left side. The initial density $n_{ini}$ is uniform within $10\lambda_0<x<30\lambda_0$, and the rest of the domain is vacuum. Two typical density values are adopted: low density with $n_{ini}=0.44n_{cr0}=2.78n_{cr1}$ ($\omega_0=1.5\omega_{p,ini}$), and high density with $n_{ini}=2.78n_{cr0}=17.4n_{cr1}$ ($\omega_0=0.6\omega_{p,ini}$). For each density value, three wave incidence cases are taken: pure pump wave with $a_0=10$ and $a_1=0$, pure LF wave with $a_0=0$ and $a_1=1$, and the mixing waves with $a_0=10$ and $a_1=1$ incident from the left vacuum.
	
	\begin{figure}[h]
		\includegraphics[width=\figsize\textwidth]{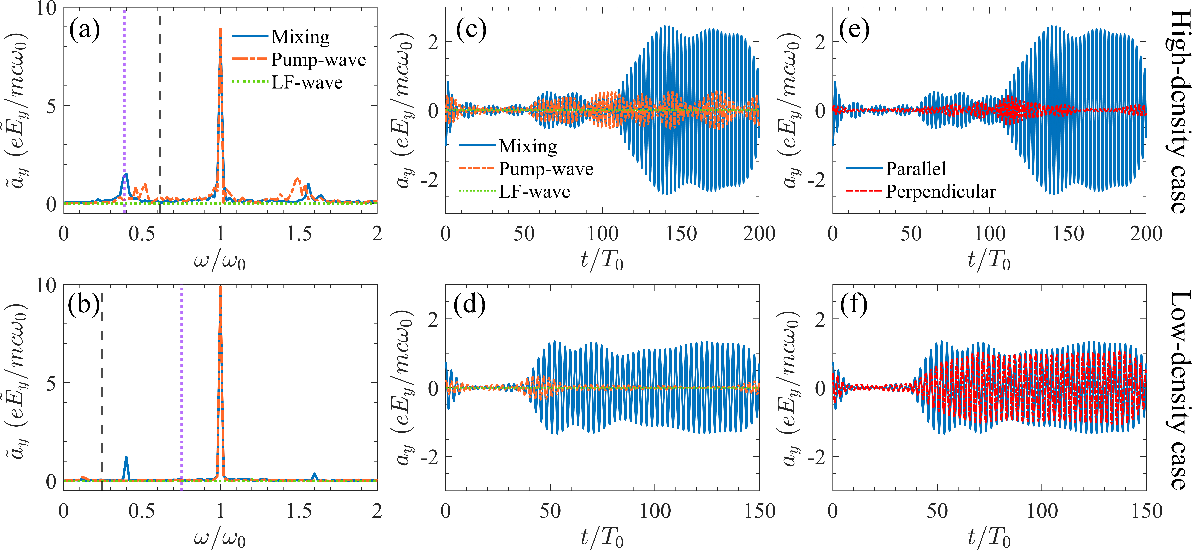}
		\caption{\label{fig:result} 
			[(a), (b)] Spectra of the laser fields collected in the right vacuum and [(c), (d)] laser field waveforms filtered with the frequency range $\omega/\omega_0\in(0.35,0.42)$, where different curves correspond to the mixing-wave, pump-wave, and LF-wave incidence cases. In plots (a) and (b), the black-dashed and purple-dotted lines represent the frequencies of $\omega_p'=\omega_p/\sqrt{\gamma_0}$ and $\omega_0-\omega_p'$. 
			[(e), (f)] Laser field waveforms filtered with $\omega/\omega_0\in(0.35,0.42)$, where the mixing waves are incident with the p-polarized LF wave and the pump wave of p-polarization (blue-solid) and s-polarization (red-dashed), respectively. The upper and lower rows correspond to high-density and low-density cases.} 
	\end{figure}

    \emph{Anomalous LF-wave transparency}.\textendash We collect the laser fields at the right vacuum after they propagate through the plasma, as illustrated in Fig. \ref{fig:result}. In the high-density plasma case with $n_{ini}=2.78n_{cr0}=17.4n_{cr1}$, the spectra of the collected fields shown in Fig. \ref{fig:result}(a) suggest that the pure pump wave can penetrate the plasma by the RT effect, but the pure LF wave cannot, as expected. Aided by the pump wave in the mixing-wave case, the LF wave of $\omega_1= 0.4\omega_0$ can penetrate the plasma, as displayed by the blue curve with a peak at  $0.4\omega_0$. This can be more clearly seen from the waveforms of the laser fields filtered within $\omega/\omega_0\in(0.35,0.42)$ in Fig. \ref{fig:result}(c). The LF wave in the mixing-wave case is much stronger than that in the pure pump-wave case, where the latter is at a low level and generated by nonlinear laser-plasma interactions [note that it is even lower in lower-density plasmas with weaker nonlinear interactions, as shown in Figs. \ref{fig:result}(d) and \ref{fig:result}(b)]. The anomalous LF-wave penetration is robust against various plasma densities and laser amplitudes, as seen in Figs. S2\textendash S4 and S16\textendash S18 in Supplemental Materials (SM) \cite{SupMat} where we vary the density within $1.11\textendash 25 n_{cr1}$ and $a_0$ within $2\textendash 20$.

    The anomalous LF-wave transparency in high-density plasmas cannot be explained by the RT effect. When we change the pump-wave polarization from parallel to perpendicular to the LF-wave polarization, the transparency disappears, as displayed in Fig. \ref{fig:result}(e). Considering the pump-wave intensity is 100 times higher than the LF wave, the relativistic modification on plasma frequency $\omega_p/\sqrt{\gamma}$ is mainly determined by the pump wave, with $\gamma \approx \gamma_0$ indepenent of the pump-wave polarization. Therefore, Fig. \ref{fig:result}(e) suggests that beyond the conventional RT effect, there is other mechanism for the anomalous LF-wave penetration. Next, we will show that it can be well explained by EIT, which is triggered by the two-wave interference, therefore, requiring the parallel polarization of the two waves.

    \emph{Model}.\textendash We develop a relativistic three-wave coupling model based on the one presented in weakly-relativistic regime \cite{Gordon2000}, where an infinite, homogeneous 1D cold fluid approximation is taken. Frequency $\omega$ is normalized by {$\omega_0$}, wave number $k$ by {$\omega_0/c$}, $t$ by {$1/\omega_0$}, $x$ by {$c/\omega_0$}, momentum ${\bf p}$ by $mc$, laser amplitude ${\bf a}$ by $m c/e$, scalar potential $\phi$ by $m c^2/e$, and electron-density perturbation $n_1=n_e-n_{ini}$ by {$n_{cr0}$}. When the Coulomb gauge is selected, ${\bf p=a}_\perp$ is invariant, and Lorentz factor $\gamma=\sqrt{1+p_x^2+a_\perp^2}$. Take all the laser waves p-polarized, i.e., ${\bf a_\perp}=a(x,t)\hat{\bf e}_y$. Take laser amplitude form of $a=\sum_i a_i\exp[\mathbbm{i}(k_i x-\omega_i t)]/2 + c.c.$ and define $\gamma_0=\sqrt{1+\sum_i a_i^2/2}$. We consider the three-wave interaction, where $|a_0|\gg|a_1|,|a_2|$ and the phase-matching conditions $\omega_{1,2}=\omega_0\mp\Delta \omega$ and $k_{1,2}=k_0\mp\Delta k$ are satisfied. Then, the wave, Poisson, and ideal fluid equations, as given by Eqs. (S1)\textendash(S4) in SM, can be simplified by extracting $\gamma_0$ even when laser amplitude $a_0\gg 1$. Taking $\omega_{p,ini}/\omega_0=\Omega$, the wave and density-perturbation equations can be expressed by
	\begin{align}
		\left(
		\pdv[2]{x}-\pdv[2]{t}-{\Omega^2}\frac{3\gamma_0^2-1}{2\gamma_0^3}
		\right)a&=
		\frac{\Omega^2}{\gamma_0}\left(
		n_1-\frac{a^2}{2\gamma_0^2}
		\right)a,
		\label{eqn:simplified wave equation}\\
		\left(
		\pdv[2]{t}+\frac{\Omega^2}{\gamma_0}
		\right)n_1=&
		\frac{1}{2\gamma_0^2}
		\pdv[2]{a}{x}.
		\label{eqn:simplified density equation}
	\end{align}

\begin{figure}[h]
		\includegraphics[width=\figsize\textwidth]{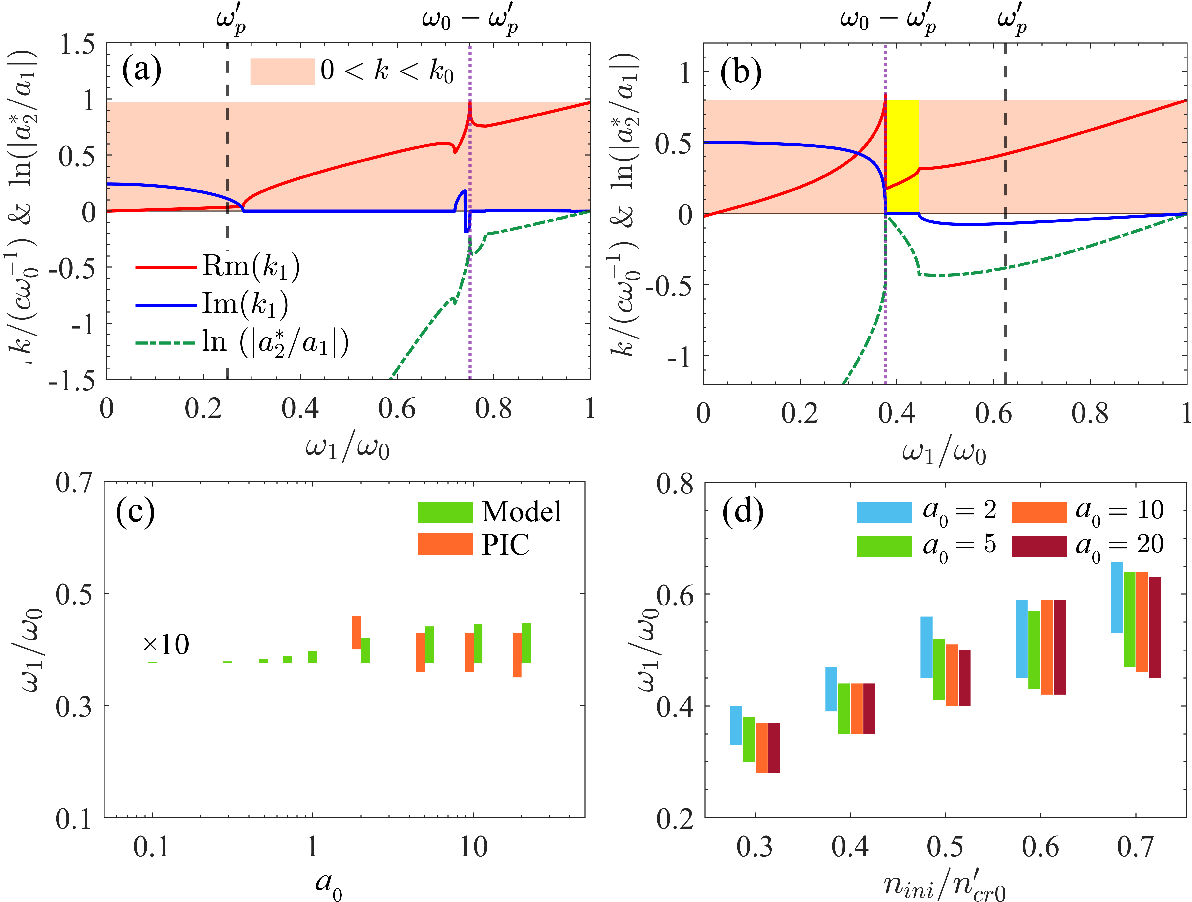}
		\caption{\label{fig:dispersion} Dispersion relationship of Stokes-dominant branches with $\ln (|a_2^*/a_1|)<0$ calculated from Eq. (\ref{eqn:dispersion relationship})  in (a) the low-density case with {$\omega_p=\omega_0/1.5$} and (b) the high-density case with {$\omega_p=\omega_0/0.6$}, where the black-dashed and purple-dotted lines represent the frequencies of $\omega_p'=\omega_p/\sqrt{\gamma_0}$ and $\omega_0-\omega_p'$. ${\rm Re}(k_1)$ (red curve) located in the orange area means forward propagation. The bright yellow area in (b) highlights the passband of $\omega_1$ within $[0.377,0.446]\omega_0$. (c) {Passband veruse $a_0$, where the green and orange bars correspond to the analytical and PIC results ($a_0\geq 2$). Initial  plasma density keeps $n_{ini}/n'_{cr0}=0.389$, corresponding to the high-density case in (b). Passband width at $a_0=0.1$ is magnified by 10 times. (d) Passband obtained by PIC results versus the density.}}
	\end{figure}

	One can find that increasing $\gamma_0$ causes decrease in both the effective plasma frequency [the 3rd term on the left-hand side of Eq. (\ref{eqn:simplified wave equation})] and nonlinear induced current (the right-hand side). The relativistically-decreased current is easier to be canceled by the beat current of pump and plasma oscillation \cite{Harris1996,Gordon2000}, lessening the requirement of EIT occurrence. In other words, the relativistic effect can broaden passband for EIT, as will be shown in Fig. \ref{fig:dispersion}(c). Taking Fourier transformation, the dispersion relationship in strongly-relativistic regime can be obtained (see complete derivation in Sec. III in SM):
	\begin{equation}
		\left[
		M_- - \frac{2\gamma_0^3}{{\Omega^2}(\gamma_0^2-1)}L_-
		\right]
		\left[
		M_+ - \frac{2\gamma_0^3}{{\Omega^2}(\gamma_0^2-1)}L_+
		\right]=M^2,
		\label{eqn:dispersion relationship}
	\end{equation}
	where
	\begin{equation}
		\begin{aligned}
			&L_\pm=({1}\pm\Delta \omega)^2-(k_0\pm\Delta k)^2-{\Omega^2}\frac{3\gamma_0^2-1}{2\gamma_0^3},\\
			&M_\pm=\frac{\Delta k^2}{\Delta \omega^2-{\Omega^2}\gamma_0^{-1}}+\frac{(2k_0\pm\Delta k)^2}{({2}\pm\Delta \omega)^2-{\Omega^2}\gamma_0^{-1}}-3,\\
			&M=\frac{\Delta k^2}{\Delta \omega^2-{\Omega^2}\gamma_0^{-1}}+\frac{2k_0^2}{{4}-{\Omega^2}\gamma_0^{-1}}-\frac{3}{2},\\
			&k_0=
			\sqrt{
				\left(
				1-{\Omega^2}\frac{1+3\gamma_0^2}{4\gamma_0^3}
				\right)\bigg/
				\frac{4-{\Omega^2}\gamma_0^{-3}}{4-{\Omega^2}\gamma_0^{-1}}}.
			\label{eqn:definitions}
		\end{aligned}
	\end{equation}

	Equations (\ref{eqn:dispersion relationship}) and (\ref{eqn:definitions}) can converge to the dispersion relationship in weakly-relativistic case presented in \cite{Gordon2000} as $\gamma_0\rightarrow 1$. For a given set of parameters $(\omega_0,\gamma_0,\Delta\omega)$, Eq. (\ref{eqn:dispersion relationship}) is a quartic equation for $\Delta k$ with four branches of solutions. By evaluating the ratio $|a_2^*/a_1|$, these solutions can be classified into Stokes and anti-Stokes dominant cases and each case has two branches. We focus on Stokes-dominant branches ($\ln |a_2^*/a_1|<0$) in which there is one representing the forward propagation of the three waves, as shown in Fig. \ref{fig:dispersion} (the complete four branches of solutions are given in Fig. S7 in SM). 
	
	\emph{Criterion and passband for stable propagation}.\textendash Stable forward propagation of both Stokes and anti-Stokes waves requires: (i) Im$(k_1)=0$; (ii) ${\rm Re}(k_1)\in(0,k_0)$. We analyze the stable propagation based on two characteristic frequencies: the effective plasma frequency $\omega_p'=\omega_p/\sqrt{\gamma_0}$ and beat-wave resonance frequency $\omega_0-\omega_p'$, where they represent the cutoff frequency for RT and a singularity of Eq. (\ref{eqn:dispersion relationship}), respectively. When $\omega_1 > \omega_p'$ ($\omega_1 < \omega_p$ always adopted), the Stokes wave can propagate by RT with the relativistic decrease of plasma frequency, corresponding to our low-density case shown in Fig. \ref{fig:dispersion}(a). In this figure, Im$(k_1)=0$ (blue curve) gives two frequency endpoints leaving a wide region for propagation, where the region lies within ($\omega_p'$, $\omega_0-\omega_p'$). Because $\omega_1< \omega_0-\omega_p'$, the current induced by the Stokes wave cannot be canceled by the beat wave of $\omega_0-\omega_p'$ \cite{Harris1996,Gordon2000}, i.e., EIT cannot be triggered. Hence, the main mechanism of the Stokes-wave propagation in this case is RT rather than EIT, supported by Fig. \ref{fig:result}(f) in which the Stokes wave (LF wave) propagation is slightly affected by the laser polarizations.
	
	While $\omega_1< \omega_p'$ and $\omega_1> \omega_0-\omega_p'$, RT does not work for the Stokes wave, meanwhile, the current induced by the Stokes wave can be canceled by the beat \cite{Harris1996,Gordon2000}, i.e., EIT can work. This is the case with the high density shown in Fig. \ref{fig:dispersion}(b). The black-dashed line representing $\omega_p'$ on the right of the purple-dotted line representing $\omega_0-\omega_p'$ (or $\omega_0 <2\omega_p'$ corresponding to the high-density case), between which there is a passband for the Stokes wave propagation, given by Im$(k_1)=0$ (blue curve). The passband of $\omega_1$ is highlighted with bright yellow and lies within $[0.377,0.446]\omega_0$, where the starting point is at $\omega_0-\omega_p'$. The obtained passband is consistent with Fig. \ref{fig:result}(a) where $\omega_1=0.4\omega_0$. In weakly-relativistic regime, our model shows that the passband is narrowed to nearly an isolated point at $\omega_1=\omega_0-\omega_p$ [{green bars in} Fig. \ref{fig:dispersion}(c) and Fig. S6 in SM]. Besides, the Stokes-wave propagation in EIT is aided by the pump wave so $\omega_0>\omega_p'$ is required for the pump wave transparent in the plasma (see more PIC simulations with different plasma densities in Figs. S2\textendash S4 in SM). To sum up, one can obtain rough criteria for EIT occurrence:
	\begin{equation}
		\omega_1>\omega_0-\frac{\omega_{p}}{\sqrt{\gamma_0}}; \quad
		\frac{2\omega_{p}}{\sqrt{\gamma_0}}
		>\omega_0>
		\frac{\omega_{p}}{\sqrt{\gamma_0}}.
		\label{eqn:criterion}
	\end{equation}
	Note that calculating Eq. (\ref{eqn:dispersion relationship}) is needed to obtain an accurater passband, as done in Figs. \ref{fig:dispersion}(b) and (c). {The calculated passband agrees with PIC results in strongly-relativistic regime [Fig. \ref{fig:dispersion}(c)]. In Fig. \ref{fig:dispersion}(d), we change plasma density and the passband exists stably and when $a_0\gg 1$, passband width and location tend to be constant at a given $n_{ini}/n_{cr0}'$, agreeing well with our analysis on the asymptotic behavior [Eqs. (S32) and (S33) in SM].}
%At the non-relativistic limit, the criterion degenerates to $\omega_1>\omega_0-\omega_p$, 
%$2\omega_p>\omega_0>\omega_p$.	
	
\begin{figure}[h]
		\includegraphics[width=\figsize\textwidth]{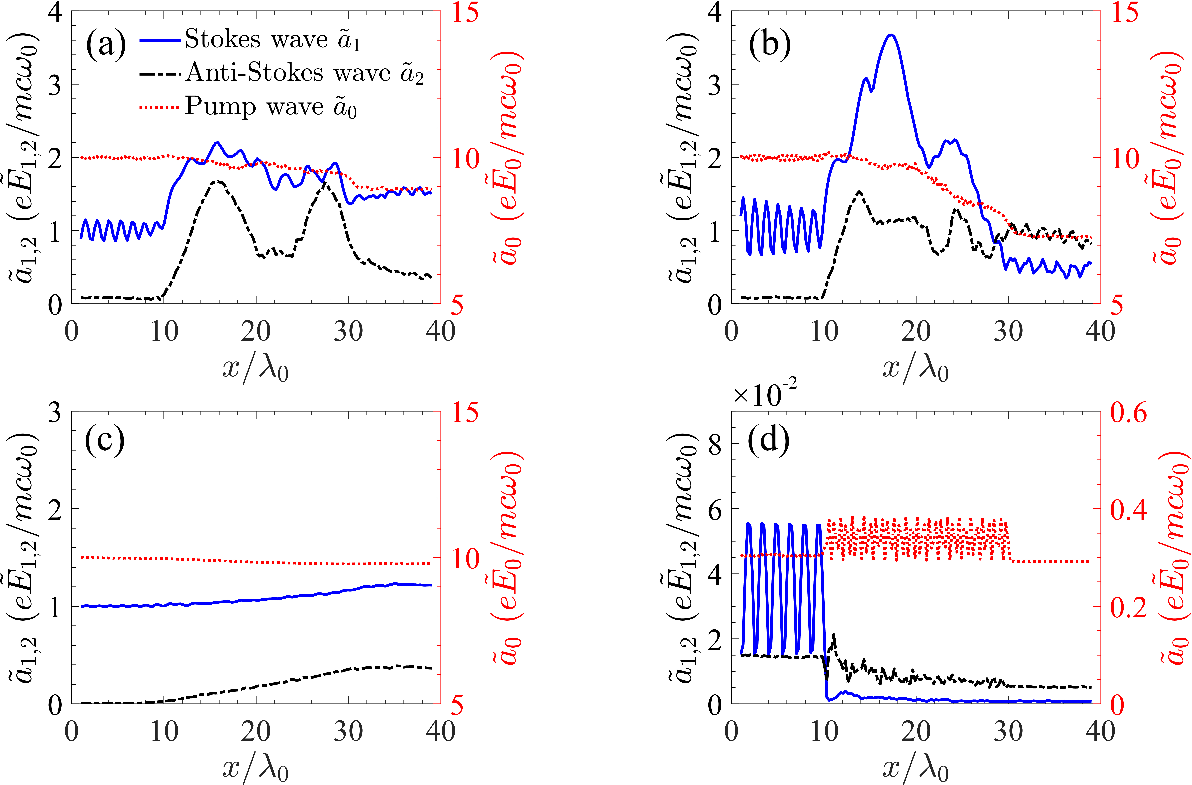}
		\caption{\label{fig: spacial evolution} Amplitudes of the Stokes and anti-Stokes waves (left-axis) as well as the pump wave (right-axis), where $a_0=10$ and $a_1=1$ are taken in (a)\textendash(c), and $a_0=0.3$ and $a_1=0.03$ are taken in (d). The plasma density $n_{ini}$ is adopted as $2.78 n_{cr0}$ ($\omega_0=0.6\omega_{p,ini}$) in (a), $3.31n_{cr0}$ ($\omega_0=0.55\omega_{p,ini}$) in (b), and $0.44 n_{cr0}$ ($\omega_0=1.5\omega_{p,ini}$) in (c) and (d). We take $\omega_1=0.4\omega_0$ in (a)\textendash(c) and $\omega_1=\omega_0/3$ in (d). Note that the plasma is located within $10\lambda_0<x<30\lambda_0$.}
	\end{figure}

	To further examine the model and investigate energy coupling among the three waves, we present spatial distributions of the wave amplitudes obtained from PIC simulations in four cases. In Fig. \ref{fig: spacial evolution}(a), the same parameters as Fig. \ref{fig:result}(a) are taken with the high-density $n_{ini}=2.78n_{cr0}$ and the LF (Stokes) wave frequency $\omega_1=0.4\omega_0$ located in the passband. Via relativistic EIT, the Stokes wave with around 2 times of the initial amplitude is transported to the right vacuum ($x>30\lambda_0$). In the plasma, energy transfer from the relativistic pump wave to the other two waves takes place, causing significant vibration and amplification of the Stokes wave and establishment of the anti-Stokes wave. When the plasma density increases [e.g., $n_{ini}=3.31 n_{cr0}$ in Fig. \ref{fig: spacial evolution}(b)], the energy transfer becomes stronger. In this case, both pump and Stokes waves experience significant energy depletion during transport, hence, a weaker Stokes signal is collected at the right vacuum. With higher density [e.g., $n_{ini}=4n_{cr0}$, see Fig. S3 in SM], stronger energy depletion occurs, even leading to the opacity of the Stokes wave. While the plasma density is relatively low [e.g., $n_{ini}=0.44 n_{cr0}$ in Fig. \ref{fig: spacial evolution}(c) and Fig. \ref{fig:result}(b)], the EIT criteria given in Eq. (\ref{eqn:criterion}) are not satisfied and therefore the energy coupling among the three waves is weak, where the Stokes amplitude appears little vibration during transport in plasmas.
	
    \emph{Quenching of EIT in weakly-relativistic regime}.\textendash It has been reported that in weakly-relativistic regime EIT cannot work in bounded plasmas \cite{Gordon2000a}. Our PIC simulation obtains the same result, as displayed in Fig. \ref{fig: spacial evolution}(d) where $a_0=0.3, a_1=0.03$ and $n_{ini}=0.44 n_{cr0}$ (i.e., $\omega_p=2\omega_0/3$). Here, we take $\omega_1=\omega_0/3$ to meet the predicted passband for Stokes-wave propagation. The pump wave penetrates through the plasma, but the Stokes wave is completely reflected on the plasma boundary $x=10\lambda_0$. On the boundary, a weak anti-Stokes wave is generated by the three-wave coupling but it does not continue to generate in the plasma with neither Stokes wave nor the subsequent three-wave coupling.  Because the anti-Stokes wave source locates at the boundary, only the forward-emitting signal (toward the plasma) attenuates gradually with propagation, while the backward-emitting one (toward the left vacuum) does not.

    Why EIT can occur in strongly-relativistic regime but not in weakly-relativistic regime? Firstly, the passband for the phase-matching condition is sufficiently wide in the former [Figs. \ref{fig:dispersion}(b)-(d)] and in the latter it narrows to nearly an isolated point $\omega_1=\omega_0-\omega_p$ (Figs. \ref{fig:dispersion}(c)). This phase-matching can be satisfied only at the beginning of the laser interaction and on the plasma boundary. Later, the plasma density is perturbed, $\omega_p$ is changed, and $\omega_1=\omega_0-\omega_p$ does not hold. This causes the three-wave coupling to occur only on the plasma boundary and EIT cannot be developed in the plasma, as shown in Fig. \ref{fig: spacial evolution}(d). In strongly-relativistic regime, the wide passband can bear perturbation in the plasma density and $\omega_p$ to a great degree.  

    Beyond the passband due to the density perturbation, the Stokes wave in strongly-relativistic regime can propagate an appreciable distance in plasmas, because the real and imaginary part of $k_1=K+\mathbbm{i}\kappa$ is comparable, which opens up opportunities to return the passband. As shown in \cite{Gordon2000a}, when the Stokes wave and pump wave reaches the boundary of the plasma, the density and nonlinear current perturbation has the form of
    \begin{equation}
    	n_1\propto a_0 a_1 e^{\mathbbm{i}(\Delta k x - \Delta\omega t)},\quad
    	j\propto |a_0|^2 a_1 e^{\mathbbm{i}(k_1 x - \omega_1 t)}.
    \end{equation}
	In weakly-relativistic regime ($|a_0|\ll 1$), the scale analysis in Eqs. (S23)\textendash(S27) in SM shows that $K$ falls rapidly while $\kappa$ soars at the growth rate $\sim O(|a_0|^{-2})$ once the frequencies of the beat and Stokes waves mismatches, resulting in $K\ll\kappa$ and $k_1\simeq \mathbbm{i}\kappa$. Therefore, the current source related to the Stokes wave $j\propto \exp(-\kappa x)$ damps rapidly and cannot penetrate into the plasma. However, in strongly-relativistic regime, the dispersion curves in Fig. \ref{fig:dispersion}(b) show that ${\rm Re}(k_1)$ on the left of beat-wave resonance line falls to zero much more gently than in weakly-relativistic regime [Figs. S6(c) and S6(d) in SM], indicating that $K\gtrsim\kappa$ even if the beat-wave frequency exceeds $\omega_1$ in a large range. Therefore, the nonlinear current and the Stokes wave can propagate in the bulk plasma before significant damping.

	\emph{Summary and Discussion}.\textendash We have found anomalous transparency of a low-frequency laser in over-dense plasmas with assistance of a relativistic pump laser. This anomalous transparency depends on the laser polarization, which can be explained by EIT rather than the well-known RT. We have developed a relativistic three-wave coupling model and presented a criterion and a frequency passband for EIT. In strongly-relativistic regime, since the relativistic effect leads to the decrease of the induced current, the passband becomes sufficiently wide, allowing stable propagation of the low-frequency laser in over-dense plasmas. EIT is sustainable over long-distance propagation and can survive considering multi-dimensional effects (Sec. VIII and IX in SM). However, the passband narrows to nearly an isolated point in weakly-relativistic regime, causing disappearance of EIT in real plasmas with boundaries, as reported in existing works. Therefore, this work could clarify the long-standing theoretic problem of EIT in plasma physics, which would inspire further investigations on multi-laser driven nonlinear processes in relativistic laser-plasma physics.
	 
On the other hand, our results can be directly applied in fast ignition or double-cone ignition (DCI) \cite{Zhang2020} of ICF. As an advanced fast ignition scheme, the idea of DCI was reported in 2020 \cite{Zhang2020}, which has been funded by a project in China since 2020. Supported by this project, ps laser beams with the total energy about 10 kJ is under construction. Besides the fundamental-frequency (1$\omega$, 1.06-micron wavelength) mode, the ps beams will have double-harmonic (2$\omega$) mode to control the generated fast-electron energy spectrum and enhance the laser transparency. The 2$\omega$ laser beams are converted from the 1$\omega$ beams and the conversion efficiency of nonlinear crystals can reach 60\textemdash70\%, and the rest energy 20\%-30\% remains as 1$\omega$ component. Whether we should retain or filter out the 1$\omega$ component for the fast-electron generation in the DCI or fast-ignition heating needes to clarify. Our recent investigation [arXiv:2311.11625] showed that mixing the higher-intensity 2$\omega$ beam with the lower-intensity 1$\omega$ component can enhance the fast-electron yield and control the spectrum, therefore, the 1$\omega$ beam should be retained. Particularly, the 2D PIC simulation illustrated that enhancement of the fast-electron yield is significant and deeper laser penetration can be observed even if the intensity of the 1$\omega$ beam is only ~$5\%$ of the 2$\omega$ beam intensity. Although the intensity of the 1$\omega$ beam (i.e., low-frequency wave) is low, it can play an important role, which cannot be explained by the relativistic transparency and laser hole boring. We think that EIT could work. To exclude laser hole boring effect, we have adopted 1D simulation and taken the low-frequency beam intensity much lower than the pump beam to distinguish EIT from relativistic transparency more clearly. 1D PIC simulation has shown that EIT arise well, which has also been verified by  2D PIC simulation (see Sec. IX in SM).

	\begin{acknowledgments}
		The authors gratefully acknowledge Dr. Shao-Jun Wang for constructive discussions. This work was supported by the Strategic Priority Research Program of Chinese Academy of Sciences (Grant Nos. XDA25050300, XDA25010300, XDA25010100), the National Key R\&D Program of China (Grant No. 2018YFA0404801), and National Natural Science Foundation of China (Grant Nos. 11827807, 11775302), and the Fundamental Research Funds for the Central Universities, the Research Funds of Renmin University of China (20XNLG01). Computational resources have been partially provided by the Physical Laboratory of High Performance Computing at Renmin University of China.
	\end{acknowledgments}
	
	%\bibliography{Reference}

\begin{thebibliography}{33}%
		\makeatletter
		\providecommand \@ifxundefined [1]{%
			\@ifx{#1\undefined}
		}%
		\providecommand \@ifnum [1]{%
			\ifnum #1\expandafter \@firstoftwo
			\else \expandafter \@secondoftwo
			\fi
		}%
		\providecommand \@ifx [1]{%
			\ifx #1\expandafter \@firstoftwo
			\else \expandafter \@secondoftwo
			\fi
		}%
		\providecommand \natexlab [1]{#1}%
		\providecommand \enquote  [1]{``#1''}%
		\providecommand \bibnamefont  [1]{#1}%
		\providecommand \bibfnamefont [1]{#1}%
		\providecommand \citenamefont [1]{#1}%
		\providecommand \href@noop [0]{\@secondoftwo}%
		\providecommand \href [0]{\begingroup \@sanitize@url \@href}%
		\providecommand \@href[1]{\@@startlink{#1}\@@href}%
		\providecommand \@@href[1]{\endgroup#1\@@endlink}%
		\providecommand \@sanitize@url [0]{\catcode `\\12\catcode `\$12\catcode
			`\&12\catcode `\#12\catcode `\^12\catcode `\_12\catcode `\%12\relax}%
		\providecommand \@@startlink[1]{}%
		\providecommand \@@endlink[0]{}%
		\providecommand \url  [0]{\begingroup\@sanitize@url \@url }%
		\providecommand \@url [1]{\endgroup\@href {#1}{\urlprefix }}%
		\providecommand \urlprefix  [0]{URL }%
		\providecommand \Eprint [0]{\href }%
		\providecommand \doibase [0]{http://dx.doi.org/}%
		\providecommand \selectlanguage [0]{\@gobble}%
		\providecommand \bibinfo  [0]{\@secondoftwo}%
		\providecommand \bibfield  [0]{\@secondoftwo}%
		\providecommand \translation [1]{[#1]}%
		\providecommand \BibitemOpen [0]{}%
		\providecommand \bibitemStop [0]{}%
		\providecommand \bibitemNoStop [0]{.\EOS\space}%
		\providecommand \EOS [0]{\spacefactor3000\relax}%
		\providecommand \BibitemShut  [1]{\csname bibitem#1\endcsname}%
		\let\auto@bib@innerbib\@empty
		%</preamble>
		\bibitem [{\citenamefont {Esarey}\ \emph {et~al.}(2009)\citenamefont {Esarey},
			\citenamefont {Schroeder},\ and\ \citenamefont {Leemans}}]{Esarey2009}%
		\BibitemOpen
		\bibfield  {author} {\bibinfo {author} {\bibfnamefont {E.}~\bibnamefont
				{Esarey}}, \bibinfo {author} {\bibfnamefont {C.~B.}\ \bibnamefont
				{Schroeder}}, \ and\ \bibinfo {author} {\bibfnamefont {W.~P.}\ \bibnamefont
				{Leemans}},\ }\href {\doibase 10.1103/RevModPhys.81.1229} {\bibfield
			{journal} {\bibinfo  {journal} {Reviews of Modern Physics}\ }\textbf
			{\bibinfo {volume} {81}},\ \bibinfo {pages} {1229} (\bibinfo {year}
			{2009})}\BibitemShut {NoStop}%
		\bibitem [{\citenamefont {Macchi}\ \emph {et~al.}(2013)\citenamefont {Macchi},
			\citenamefont {Borghesi},\ and\ \citenamefont {Passoni}}]{Macchi2013}%
		\BibitemOpen
		\bibfield  {author} {\bibinfo {author} {\bibfnamefont {A.}~\bibnamefont
				{Macchi}}, \bibinfo {author} {\bibfnamefont {M.}~\bibnamefont {Borghesi}}, \
			and\ \bibinfo {author} {\bibfnamefont {M.}~\bibnamefont {Passoni}},\ }\href
		{\doibase 10.1103/RevModPhys.85.751} {\bibfield  {journal} {\bibinfo
				{journal} {Reviews of Modern Physics}\ }\textbf {\bibinfo {volume} {85}},\
			\bibinfo {pages} {751} (\bibinfo {year} {2013})},\ \Eprint
		{http://arxiv.org/abs/1302.1775} {arXiv:1302.1775} \BibitemShut {NoStop}%
		\bibitem [{\citenamefont {Armstrong}\ \emph {et~al.}(2019)\citenamefont
			{Armstrong}, \citenamefont {Brenner}, \citenamefont {Zemaityte},
			\citenamefont {Scott}, \citenamefont {Rusby}, \citenamefont {Liao},
			\citenamefont {Liu}, \citenamefont {Li}, \citenamefont {Zhang}, \citenamefont
			{Zhang}, \citenamefont {Zhu}, \citenamefont {Bradford}, \citenamefont
			{Woolsey}, \citenamefont {Oliveira}, \citenamefont {Spindloe}, \citenamefont
			{Wang}, \citenamefont {McKenna},\ and\ \citenamefont
			{Neely}}]{Armstrong2019}%
		\BibitemOpen
		\bibfield  {author} {\bibinfo {author} {\bibfnamefont {C.~D.}\ \bibnamefont
				{Armstrong}}, \bibinfo {author} {\bibfnamefont {C.~M.}\ \bibnamefont
				{Brenner}}, \bibinfo {author} {\bibfnamefont {E.}~\bibnamefont {Zemaityte}},
			\bibinfo {author} {\bibfnamefont {G.~G.}\ \bibnamefont {Scott}}, \bibinfo
			{author} {\bibfnamefont {D.~R.}\ \bibnamefont {Rusby}}, \bibinfo {author}
			{\bibfnamefont {G.}~\bibnamefont {Liao}}, \bibinfo {author} {\bibfnamefont
				{H.}~\bibnamefont {Liu}}, \bibinfo {author} {\bibfnamefont {Y.}~\bibnamefont
				{Li}}, \bibinfo {author} {\bibfnamefont {Z.}~\bibnamefont {Zhang}}, \bibinfo
			{author} {\bibfnamefont {Y.}~\bibnamefont {Zhang}}, \bibinfo {author}
			{\bibfnamefont {B.}~\bibnamefont {Zhu}}, \bibinfo {author} {\bibfnamefont
				{P.}~\bibnamefont {Bradford}}, \bibinfo {author} {\bibfnamefont {N.~C.}\
				\bibnamefont {Woolsey}}, \bibinfo {author} {\bibfnamefont {P.}~\bibnamefont
				{Oliveira}}, \bibinfo {author} {\bibfnamefont {C.}~\bibnamefont {Spindloe}},
			\bibinfo {author} {\bibfnamefont {W.}~\bibnamefont {Wang}}, \bibinfo {author}
			{\bibfnamefont {P.}~\bibnamefont {McKenna}}, \ and\ \bibinfo {author}
			{\bibfnamefont {D.}~\bibnamefont {Neely}},\ }\href {\doibase
			10.1088/1361-6587/aaf596} {\bibfield  {journal} {\bibinfo  {journal} {Plasma
					Physics and Controlled Fusion}\ }\textbf {\bibinfo {volume} {61}},\ \bibinfo
			{pages} {034001} (\bibinfo {year} {2019})}\BibitemShut {NoStop}%
		\bibitem [{\citenamefont {Schoenlein}\ \emph {et~al.}(2019)\citenamefont
			{Schoenlein}, \citenamefont {Elsaesser}, \citenamefont {Holldack},
			\citenamefont {Huang}, \citenamefont {Kapteyn}, \citenamefont {Murnane},\
			and\ \citenamefont {Woerner}}]{Schoenlein2019}%
		\BibitemOpen
		\bibfield  {author} {\bibinfo {author} {\bibfnamefont {R.}~\bibnamefont
				{Schoenlein}}, \bibinfo {author} {\bibfnamefont {T.}~\bibnamefont
				{Elsaesser}}, \bibinfo {author} {\bibfnamefont {K.}~\bibnamefont {Holldack}},
			\bibinfo {author} {\bibfnamefont {Z.}~\bibnamefont {Huang}}, \bibinfo
			{author} {\bibfnamefont {H.}~\bibnamefont {Kapteyn}}, \bibinfo {author}
			{\bibfnamefont {M.}~\bibnamefont {Murnane}}, \ and\ \bibinfo {author}
			{\bibfnamefont {M.}~\bibnamefont {Woerner}},\ }\href {\doibase
			10.1098/rsta.2018.0384} {\bibfield  {journal} {\bibinfo  {journal}
				{Philosophical Transactions of the Royal Society A: Mathematical, Physical
					and Engineering Sciences}\ }\textbf {\bibinfo {volume} {377}},\ \bibinfo
			{pages} {20180384} (\bibinfo {year} {2019})}\BibitemShut {NoStop}%
		\bibitem [{\citenamefont {Rosmej}\ \emph {et~al.}(2019)\citenamefont {Rosmej},
			\citenamefont {Andreev}, \citenamefont {Zaehter}, \citenamefont {Zahn},
			\citenamefont {Christ}, \citenamefont {Borm}, \citenamefont {Radon},
			\citenamefont {Sokolov}, \citenamefont {Pugachev}, \citenamefont {Khaghani},
			\citenamefont {Horst}, \citenamefont {Borisenko}, \citenamefont {Sklizkov},\
			and\ \citenamefont {Pimenov}}]{Rosmej2019}%
		\BibitemOpen
		\bibfield  {author} {\bibinfo {author} {\bibfnamefont {O.~N.}\ \bibnamefont
				{Rosmej}}, \bibinfo {author} {\bibfnamefont {N.~E.}\ \bibnamefont {Andreev}},
			\bibinfo {author} {\bibfnamefont {S.}~\bibnamefont {Zaehter}}, \bibinfo
			{author} {\bibfnamefont {N.}~\bibnamefont {Zahn}}, \bibinfo {author}
			{\bibfnamefont {P.}~\bibnamefont {Christ}}, \bibinfo {author} {\bibfnamefont
				{B.}~\bibnamefont {Borm}}, \bibinfo {author} {\bibfnamefont {T.}~\bibnamefont
				{Radon}}, \bibinfo {author} {\bibfnamefont {A.}~\bibnamefont {Sokolov}},
			\bibinfo {author} {\bibfnamefont {L.~P.}\ \bibnamefont {Pugachev}}, \bibinfo
			{author} {\bibfnamefont {D.}~\bibnamefont {Khaghani}}, \bibinfo {author}
			{\bibfnamefont {F.}~\bibnamefont {Horst}}, \bibinfo {author} {\bibfnamefont
				{N.~G.}\ \bibnamefont {Borisenko}}, \bibinfo {author} {\bibfnamefont
				{G.}~\bibnamefont {Sklizkov}}, \ and\ \bibinfo {author} {\bibfnamefont
				{V.~G.}\ \bibnamefont {Pimenov}},\ }\href {\doibase 10.1088/1367-2630/ab1047}
		{\bibfield  {journal} {\bibinfo  {journal} {New Journal of Physics}\ }\textbf
			{\bibinfo {volume} {21}},\ \bibinfo {pages} {043044} (\bibinfo {year}
			{2019})}\BibitemShut {NoStop}%
		\bibitem [{\citenamefont {Chopineau}\ \emph {et~al.}(2019)\citenamefont
			{Chopineau}, \citenamefont {Leblanc}, \citenamefont {Blaclard}, \citenamefont
			{Denoeud}, \citenamefont {Th{\'{e}}venet}, \citenamefont {Vay}, \citenamefont
			{Bonnaud}, \citenamefont {Martin}, \citenamefont {Vincenti},\ and\
			\citenamefont {Qu{\'{e}}r{\'{e}}}}]{Chopineau2019}%
		\BibitemOpen
		\bibfield  {author} {\bibinfo {author} {\bibfnamefont {L.}~\bibnamefont
				{Chopineau}}, \bibinfo {author} {\bibfnamefont {A.}~\bibnamefont {Leblanc}},
			\bibinfo {author} {\bibfnamefont {G.}~\bibnamefont {Blaclard}}, \bibinfo
			{author} {\bibfnamefont {A.}~\bibnamefont {Denoeud}}, \bibinfo {author}
			{\bibfnamefont {M.}~\bibnamefont {Th{\'{e}}venet}}, \bibinfo {author}
			{\bibfnamefont {J.-L.}\ \bibnamefont {Vay}}, \bibinfo {author} {\bibfnamefont
				{G.}~\bibnamefont {Bonnaud}}, \bibinfo {author} {\bibfnamefont
				{P.}~\bibnamefont {Martin}}, \bibinfo {author} {\bibfnamefont
				{H.}~\bibnamefont {Vincenti}}, \ and\ \bibinfo {author} {\bibfnamefont
				{F.}~\bibnamefont {Qu{\'{e}}r{\'{e}}}},\ }\href {\doibase
			10.1103/PhysRevX.9.011050} {\bibfield  {journal} {\bibinfo  {journal}
				{Physical Review X}\ }\textbf {\bibinfo {volume} {9}},\ \bibinfo {pages}
			{011050} (\bibinfo {year} {2019})},\ \Eprint
		{http://arxiv.org/abs/1809.03903} {arXiv:1809.03903} \BibitemShut {NoStop}%
		\bibitem [{\citenamefont {Nuckolls}\ \emph {et~al.}(1972)\citenamefont
			{Nuckolls}, \citenamefont {Wood}, \citenamefont {Thiessen},\ and\
			\citenamefont {Zimmerman}}]{Nuckolls1972}%
		\BibitemOpen
		\bibfield  {author} {\bibinfo {author} {\bibfnamefont {J.}~\bibnamefont
				{Nuckolls}}, \bibinfo {author} {\bibfnamefont {L.}~\bibnamefont {Wood}},
			\bibinfo {author} {\bibfnamefont {A.}~\bibnamefont {Thiessen}}, \ and\
			\bibinfo {author} {\bibfnamefont {G.}~\bibnamefont {Zimmerman}},\ }\href
		{\doibase 10.1038/239139a0} {\bibfield  {journal} {\bibinfo  {journal}
				{Nature}\ }\textbf {\bibinfo {volume} {239}},\ \bibinfo {pages} {139}
			(\bibinfo {year} {1972})}\BibitemShut {NoStop}%
		\bibitem [{\citenamefont {Kodama}\ \emph {et~al.}(2001)\citenamefont {Kodama},
			\citenamefont {Norreys}, \citenamefont {Mima}, \citenamefont {Dangor},
			\citenamefont {Evans}, \citenamefont {Fujita}, \citenamefont {Kitagawa},
			\citenamefont {Krushelnick}, \citenamefont {Miyakoshi}, \citenamefont
			{Miyanaga}, \citenamefont {Norimatsu}, \citenamefont {Rose}, \citenamefont
			{Shozaki}, \citenamefont {Shigemori}, \citenamefont {Sunahara}, \citenamefont
			{Tampo}, \citenamefont {Tanaka}, \citenamefont {Toyama}, \citenamefont
			{Yamanaka},\ and\ \citenamefont {Zepf}}]{Kodama2001}%
		\BibitemOpen
		\bibfield  {author} {\bibinfo {author} {\bibfnamefont {R.}~\bibnamefont
				{Kodama}}, \bibinfo {author} {\bibfnamefont {P.~A.}\ \bibnamefont {Norreys}},
			\bibinfo {author} {\bibfnamefont {K.}~\bibnamefont {Mima}}, \bibinfo {author}
			{\bibfnamefont {A.~E.}\ \bibnamefont {Dangor}}, \bibinfo {author}
			{\bibfnamefont {R.~G.}\ \bibnamefont {Evans}}, \bibinfo {author}
			{\bibfnamefont {H.}~\bibnamefont {Fujita}}, \bibinfo {author} {\bibfnamefont
				{Y.}~\bibnamefont {Kitagawa}}, \bibinfo {author} {\bibfnamefont
				{K.}~\bibnamefont {Krushelnick}}, \bibinfo {author} {\bibfnamefont
				{T.}~\bibnamefont {Miyakoshi}}, \bibinfo {author} {\bibfnamefont
				{N.}~\bibnamefont {Miyanaga}}, \bibinfo {author} {\bibfnamefont
				{T.}~\bibnamefont {Norimatsu}}, \bibinfo {author} {\bibfnamefont {S.~J.}\
				\bibnamefont {Rose}}, \bibinfo {author} {\bibfnamefont {T.}~\bibnamefont
				{Shozaki}}, \bibinfo {author} {\bibfnamefont {K.}~\bibnamefont {Shigemori}},
			\bibinfo {author} {\bibfnamefont {A.}~\bibnamefont {Sunahara}}, \bibinfo
			{author} {\bibfnamefont {M.}~\bibnamefont {Tampo}}, \bibinfo {author}
			{\bibfnamefont {K.~A.}\ \bibnamefont {Tanaka}}, \bibinfo {author}
			{\bibfnamefont {Y.}~\bibnamefont {Toyama}}, \bibinfo {author} {\bibfnamefont
				{T.}~\bibnamefont {Yamanaka}}, \ and\ \bibinfo {author} {\bibfnamefont
				{M.}~\bibnamefont {Zepf}},\ }\href {\doibase 10.1038/35090525} {\bibfield
			{journal} {\bibinfo  {journal} {Nature}\ }\textbf {\bibinfo {volume} {412}},\
			\bibinfo {pages} {798} (\bibinfo {year} {2001})}\BibitemShut {NoStop}%
		\bibitem [{\citenamefont {Zhang}\ \emph {et~al.}(2020)\citenamefont {Zhang},
			\citenamefont {Wang}, \citenamefont {Yang}, \citenamefont {Wu}, \citenamefont
			{Ma}, \citenamefont {Jiao}, \citenamefont {Zhang}, \citenamefont {Wu},
			\citenamefont {Yuan}, \citenamefont {Li},\ and\ \citenamefont
			{Zhu}}]{Zhang2020}%
		\BibitemOpen
		\bibfield  {author} {\bibinfo {author} {\bibfnamefont {J.}~\bibnamefont
				{Zhang}}, \bibinfo {author} {\bibfnamefont {W.~M.}\ \bibnamefont {Wang}},
			\bibinfo {author} {\bibfnamefont {X.~H.}\ \bibnamefont {Yang}}, \bibinfo
			{author} {\bibfnamefont {D.}~\bibnamefont {Wu}}, \bibinfo {author}
			{\bibfnamefont {Y.~Y.}\ \bibnamefont {Ma}}, \bibinfo {author} {\bibfnamefont
				{J.~L.}\ \bibnamefont {Jiao}}, \bibinfo {author} {\bibfnamefont
				{Z.}~\bibnamefont {Zhang}}, \bibinfo {author} {\bibfnamefont {F.~Y.}\
				\bibnamefont {Wu}}, \bibinfo {author} {\bibfnamefont {X.~H.}\ \bibnamefont
				{Yuan}}, \bibinfo {author} {\bibfnamefont {Y.~T.}\ \bibnamefont {Li}}, \ and\
			\bibinfo {author} {\bibfnamefont {J.~Q.}\ \bibnamefont {Zhu}},\ }\href
		{\doibase 10.1098/rsta.2020.0015} {\bibfield  {journal} {\bibinfo  {journal}
				{Philosophical transactions. Series A, Mathematical, physical, and
					engineering sciences}\ }\textbf {\bibinfo {volume} {378}},\ \bibinfo {pages}
			{20200015} (\bibinfo {year} {2020})}\BibitemShut {NoStop}%
		\bibitem [{\citenamefont {Miquel}\ and\ \citenamefont
			{Prene}(2019)}]{Miquel2019}%
		\BibitemOpen
		\bibfield  {author} {\bibinfo {author} {\bibfnamefont {J.~L.}\ \bibnamefont
				{Miquel}}\ and\ \bibinfo {author} {\bibfnamefont {E.}~\bibnamefont {Prene}},\
		}\href {\doibase 10.1088/1741-4326/aac343} {\bibfield  {journal} {\bibinfo
				{journal} {Nuclear Fusion}\ }\textbf {\bibinfo {volume} {59}} (\bibinfo
			{year} {2019}),\ 10.1088/1741-4326/aac343}\BibitemShut {NoStop}%
		\bibitem [{\citenamefont {Meyerhofer}\ \emph {et~al.}(2010)\citenamefont
			{Meyerhofer}, \citenamefont {Bromage}, \citenamefont {Dorrer}, \citenamefont
			{Kelly}, \citenamefont {Kruschwitz}, \citenamefont {Loucks}, \citenamefont
			{McCrory}, \citenamefont {Morse}, \citenamefont {Myatt}, \citenamefont
			{Nilson}, \citenamefont {Qiao}, \citenamefont {Sangster}, \citenamefont
			{Stoeckl}, \citenamefont {Waxer},\ and\ \citenamefont
			{Zuegel}}]{Meyerhofer2010}%
		\BibitemOpen
		\bibfield  {author} {\bibinfo {author} {\bibfnamefont {D.~D.}\ \bibnamefont
				{Meyerhofer}}, \bibinfo {author} {\bibfnamefont {J.}~\bibnamefont {Bromage}},
			\bibinfo {author} {\bibfnamefont {C.}~\bibnamefont {Dorrer}}, \bibinfo
			{author} {\bibfnamefont {J.~H.}\ \bibnamefont {Kelly}}, \bibinfo {author}
			{\bibfnamefont {B.~E.}\ \bibnamefont {Kruschwitz}}, \bibinfo {author}
			{\bibfnamefont {S.~J.}\ \bibnamefont {Loucks}}, \bibinfo {author}
			{\bibfnamefont {R.~L.}\ \bibnamefont {McCrory}}, \bibinfo {author}
			{\bibfnamefont {S.~F.~B.}\ \bibnamefont {Morse}}, \bibinfo {author}
			{\bibfnamefont {J.~F.}\ \bibnamefont {Myatt}}, \bibinfo {author}
			{\bibfnamefont {P.~M.}\ \bibnamefont {Nilson}}, \bibinfo {author}
			{\bibfnamefont {J.}~\bibnamefont {Qiao}}, \bibinfo {author} {\bibfnamefont
				{T.~C.}\ \bibnamefont {Sangster}}, \bibinfo {author} {\bibfnamefont
				{C.}~\bibnamefont {Stoeckl}}, \bibinfo {author} {\bibfnamefont {L.~J.}\
				\bibnamefont {Waxer}}, \ and\ \bibinfo {author} {\bibfnamefont {J.~D.}\
				\bibnamefont {Zuegel}},\ }\href {\doibase 10.1088/1742-6596/244/3/032010}
		{\bibfield  {journal} {\bibinfo  {journal} {Journal of Physics: Conference
					Series}\ }\textbf {\bibinfo {volume} {244}},\ \bibinfo {pages} {032010}
			(\bibinfo {year} {2010})}\BibitemShut {NoStop}%
		\bibitem [{\citenamefont {Jiao}\ \emph {et~al.}(2018)\citenamefont {Jiao},
			\citenamefont {Shao}, \citenamefont {Zhao}, \citenamefont {Wu}, \citenamefont
			{Ji}, \citenamefont {Wang}, \citenamefont {Xia}, \citenamefont {Liu},
			\citenamefont {Zhou}, \citenamefont {Ju}, \citenamefont {Cai}, \citenamefont
			{Ye}, \citenamefont {Qiao}, \citenamefont {Hua}, \citenamefont {Li},
			\citenamefont {Pan}, \citenamefont {Ren}, \citenamefont {Sun}, \citenamefont
			{Zhu},\ and\ \citenamefont {Lin}}]{Jiao2018}%
		\BibitemOpen
		\bibfield  {author} {\bibinfo {author} {\bibfnamefont {Z.}~\bibnamefont
				{Jiao}}, \bibinfo {author} {\bibfnamefont {P.}~\bibnamefont {Shao}}, \bibinfo
			{author} {\bibfnamefont {D.}~\bibnamefont {Zhao}}, \bibinfo {author}
			{\bibfnamefont {R.}~\bibnamefont {Wu}}, \bibinfo {author} {\bibfnamefont
				{L.}~\bibnamefont {Ji}}, \bibinfo {author} {\bibfnamefont {L.}~\bibnamefont
				{Wang}}, \bibinfo {author} {\bibfnamefont {L.}~\bibnamefont {Xia}}, \bibinfo
			{author} {\bibfnamefont {D.}~\bibnamefont {Liu}}, \bibinfo {author}
			{\bibfnamefont {Y.}~\bibnamefont {Zhou}}, \bibinfo {author} {\bibfnamefont
				{L.}~\bibnamefont {Ju}}, \bibinfo {author} {\bibfnamefont {Z.}~\bibnamefont
				{Cai}}, \bibinfo {author} {\bibfnamefont {Q.}~\bibnamefont {Ye}}, \bibinfo
			{author} {\bibfnamefont {Z.}~\bibnamefont {Qiao}}, \bibinfo {author}
			{\bibfnamefont {N.}~\bibnamefont {Hua}}, \bibinfo {author} {\bibfnamefont
				{Q.}~\bibnamefont {Li}}, \bibinfo {author} {\bibfnamefont {W.}~\bibnamefont
				{Pan}}, \bibinfo {author} {\bibfnamefont {L.}~\bibnamefont {Ren}}, \bibinfo
			{author} {\bibfnamefont {M.}~\bibnamefont {Sun}}, \bibinfo {author}
			{\bibfnamefont {J.}~\bibnamefont {Zhu}}, \ and\ \bibinfo {author}
			{\bibfnamefont {Z.}~\bibnamefont {Lin}},\ }\href {\doibase
			10.1017/hpl.2018.8} {\bibfield  {journal} {\bibinfo  {journal} {High Power
					Laser Science and Engineering}\ }\textbf {\bibinfo {volume} {6}},\ \bibinfo
			{pages} {1} (\bibinfo {year} {2018})}\BibitemShut {NoStop}%
		\bibitem [{\citenamefont {Kaw}\ and\ \citenamefont {Dawson}(1970)}]{Kaw1970}%
		\BibitemOpen
		\bibfield  {author} {\bibinfo {author} {\bibfnamefont {P.}~\bibnamefont
				{Kaw}}\ and\ \bibinfo {author} {\bibfnamefont {J.}~\bibnamefont {Dawson}},\
		}\href {\doibase 10.1063/1.1692942} {\bibfield  {journal} {\bibinfo
				{journal} {Physics of Fluids}\ }\textbf {\bibinfo {volume} {13}},\ \bibinfo
			{pages} {472} (\bibinfo {year} {1970})}\BibitemShut {NoStop}%
		\bibitem [{\citenamefont {Palaniyappan}\ \emph {et~al.}(2012)\citenamefont
			{Palaniyappan}, \citenamefont {Hegelich}, \citenamefont {Wu}, \citenamefont
			{Jung}, \citenamefont {Gautier}, \citenamefont {Yin}, \citenamefont
			{Albright}, \citenamefont {Johnson}, \citenamefont {Shimada}, \citenamefont
			{Letzring}, \citenamefont {Offermann}, \citenamefont {Ren}, \citenamefont
			{Huang}, \citenamefont {H{\"{o}}rlein}, \citenamefont {Dromey}, \citenamefont
			{Fernandez},\ and\ \citenamefont {Shah}}]{Palaniyappan2012}%
		\BibitemOpen
		\bibfield  {author} {\bibinfo {author} {\bibfnamefont {S.}~\bibnamefont
				{Palaniyappan}}, \bibinfo {author} {\bibfnamefont {B.~M.}\ \bibnamefont
				{Hegelich}}, \bibinfo {author} {\bibfnamefont {H.~C.}\ \bibnamefont {Wu}},
			\bibinfo {author} {\bibfnamefont {D.}~\bibnamefont {Jung}}, \bibinfo {author}
			{\bibfnamefont {D.~C.}\ \bibnamefont {Gautier}}, \bibinfo {author}
			{\bibfnamefont {L.}~\bibnamefont {Yin}}, \bibinfo {author} {\bibfnamefont
				{B.~J.}\ \bibnamefont {Albright}}, \bibinfo {author} {\bibfnamefont {R.~P.}\
				\bibnamefont {Johnson}}, \bibinfo {author} {\bibfnamefont {T.}~\bibnamefont
				{Shimada}}, \bibinfo {author} {\bibfnamefont {S.}~\bibnamefont {Letzring}},
			\bibinfo {author} {\bibfnamefont {D.~T.}\ \bibnamefont {Offermann}}, \bibinfo
			{author} {\bibfnamefont {J.}~\bibnamefont {Ren}}, \bibinfo {author}
			{\bibfnamefont {C.}~\bibnamefont {Huang}}, \bibinfo {author} {\bibfnamefont
				{R.}~\bibnamefont {H{\"{o}}rlein}}, \bibinfo {author} {\bibfnamefont
				{B.}~\bibnamefont {Dromey}}, \bibinfo {author} {\bibfnamefont {J.~C.}\
				\bibnamefont {Fernandez}}, \ and\ \bibinfo {author} {\bibfnamefont {R.~C.}\
				\bibnamefont {Shah}},\ }\href {\doibase 10.1038/nphys2390} {\bibfield
			{journal} {\bibinfo  {journal} {Nature Physics}\ }\textbf {\bibinfo {volume}
				{8}},\ \bibinfo {pages} {763} (\bibinfo {year} {2012})}\BibitemShut {NoStop}%
		\bibitem [{\citenamefont {Willingale}\ \emph {et~al.}(2009)\citenamefont
			{Willingale}, \citenamefont {Nagel}, \citenamefont {Thomas}, \citenamefont
			{Bellei}, \citenamefont {Clarke}, \citenamefont {Dangor}, \citenamefont
			{Heathcote}, \citenamefont {Kaluza}, \citenamefont {Kamperidis},
			\citenamefont {Kneip}, \citenamefont {Krushelnick}, \citenamefont {Lopes},
			\citenamefont {Mangles}, \citenamefont {Nazarov}, \citenamefont {Nilson},\
			and\ \citenamefont {Najmudin}}]{Willingale2009}%
		\BibitemOpen
		\bibfield  {author} {\bibinfo {author} {\bibfnamefont {L.}~\bibnamefont
				{Willingale}}, \bibinfo {author} {\bibfnamefont {S.~R.}\ \bibnamefont
				{Nagel}}, \bibinfo {author} {\bibfnamefont {A.~G.~R.}\ \bibnamefont
				{Thomas}}, \bibinfo {author} {\bibfnamefont {C.}~\bibnamefont {Bellei}},
			\bibinfo {author} {\bibfnamefont {R.~J.}\ \bibnamefont {Clarke}}, \bibinfo
			{author} {\bibfnamefont {A.~E.}\ \bibnamefont {Dangor}}, \bibinfo {author}
			{\bibfnamefont {R.}~\bibnamefont {Heathcote}}, \bibinfo {author}
			{\bibfnamefont {M.~C.}\ \bibnamefont {Kaluza}}, \bibinfo {author}
			{\bibfnamefont {C.}~\bibnamefont {Kamperidis}}, \bibinfo {author}
			{\bibfnamefont {S.}~\bibnamefont {Kneip}}, \bibinfo {author} {\bibfnamefont
				{K.}~\bibnamefont {Krushelnick}}, \bibinfo {author} {\bibfnamefont
				{N.}~\bibnamefont {Lopes}}, \bibinfo {author} {\bibfnamefont {S.~P.~D.}\
				\bibnamefont {Mangles}}, \bibinfo {author} {\bibfnamefont {W.}~\bibnamefont
				{Nazarov}}, \bibinfo {author} {\bibfnamefont {P.~M.}\ \bibnamefont {Nilson}},
			\ and\ \bibinfo {author} {\bibfnamefont {Z.}~\bibnamefont {Najmudin}},\
		}\href {\doibase 10.1103/PhysRevLett.102.125002} {\bibfield  {journal}
			{\bibinfo  {journal} {Physical Review Letters}\ }\textbf {\bibinfo {volume}
				{102}},\ \bibinfo {pages} {125002} (\bibinfo {year} {2009})}\BibitemShut
		{NoStop}%
		\bibitem [{\citenamefont {Weng}\ \emph {et~al.}(2012)\citenamefont {Weng},
			\citenamefont {Mulser},\ and\ \citenamefont {Sheng}}]{Weng2012}%
		\BibitemOpen
		\bibfield  {author} {\bibinfo {author} {\bibfnamefont {S.~M.}\ \bibnamefont
				{Weng}}, \bibinfo {author} {\bibfnamefont {P.}~\bibnamefont {Mulser}}, \ and\
			\bibinfo {author} {\bibfnamefont {Z.~M.}\ \bibnamefont {Sheng}},\ }\href
		{\doibase 10.1063/1.3680638} {\bibfield  {journal} {\bibinfo  {journal}
				{Physics of Plasmas}\ }\textbf {\bibinfo {volume} {19}} (\bibinfo {year}
			{2012}),\ 10.1063/1.3680638}\BibitemShut {NoStop}%
		\bibitem [{\citenamefont {Frazer}\ \emph {et~al.}(2020)\citenamefont {Frazer},
			\citenamefont {Wilson}, \citenamefont {King}, \citenamefont {Butler},
			\citenamefont {Carroll}, \citenamefont {Duff}, \citenamefont {Higginson},
			\citenamefont {Jarrett}, \citenamefont {Davidson}, \citenamefont {Armstrong},
			\citenamefont {Liu}, \citenamefont {Neely}, \citenamefont {Gray},\ and\
			\citenamefont {McKenna}}]{Frazer2020}%
		\BibitemOpen
		\bibfield  {author} {\bibinfo {author} {\bibfnamefont {T.~P.}\ \bibnamefont
				{Frazer}}, \bibinfo {author} {\bibfnamefont {R.}~\bibnamefont {Wilson}},
			\bibinfo {author} {\bibfnamefont {M.}~\bibnamefont {King}}, \bibinfo {author}
			{\bibfnamefont {N.~M.~H.}\ \bibnamefont {Butler}}, \bibinfo {author}
			{\bibfnamefont {D.~C.}\ \bibnamefont {Carroll}}, \bibinfo {author}
			{\bibfnamefont {M.~J.}\ \bibnamefont {Duff}}, \bibinfo {author}
			{\bibfnamefont {A.}~\bibnamefont {Higginson}}, \bibinfo {author}
			{\bibfnamefont {J.}~\bibnamefont {Jarrett}}, \bibinfo {author} {\bibfnamefont
				{Z.~E.}\ \bibnamefont {Davidson}}, \bibinfo {author} {\bibfnamefont
				{C.}~\bibnamefont {Armstrong}}, \bibinfo {author} {\bibfnamefont
				{H.}~\bibnamefont {Liu}}, \bibinfo {author} {\bibfnamefont {D.}~\bibnamefont
				{Neely}}, \bibinfo {author} {\bibfnamefont {R.~J.}\ \bibnamefont {Gray}}, \
			and\ \bibinfo {author} {\bibfnamefont {P.}~\bibnamefont {McKenna}},\ }\href
		{\doibase 10.1103/PhysRevResearch.2.042015} {\bibfield  {journal} {\bibinfo
				{journal} {Physical Review Research}\ }\textbf {\bibinfo {volume} {2}},\
			\bibinfo {pages} {042015(R)} (\bibinfo {year} {2020})}\BibitemShut {NoStop}%
		\bibitem [{\citenamefont {Stark}\ \emph {et~al.}(2015)\citenamefont {Stark},
			\citenamefont {Bhattacharjee}, \citenamefont {Arefiev}, \citenamefont
			{Toncian}, \citenamefont {Hazeltine},\ and\ \citenamefont
			{Mahajan}}]{Stark2015}%
		\BibitemOpen
		\bibfield  {author} {\bibinfo {author} {\bibfnamefont {D.~J.}\ \bibnamefont
				{Stark}}, \bibinfo {author} {\bibfnamefont {C.}~\bibnamefont
				{Bhattacharjee}}, \bibinfo {author} {\bibfnamefont {A.~V.}\ \bibnamefont
				{Arefiev}}, \bibinfo {author} {\bibfnamefont {T.}~\bibnamefont {Toncian}},
			\bibinfo {author} {\bibfnamefont {R.~D.}\ \bibnamefont {Hazeltine}}, \ and\
			\bibinfo {author} {\bibfnamefont {S.~M.}\ \bibnamefont {Mahajan}},\ }\href
		{\doibase 10.1103/PhysRevLett.115.025002} {\bibfield  {journal} {\bibinfo
				{journal} {Physical Review Letters}\ }\textbf {\bibinfo {volume} {115}},\
			\bibinfo {pages} {025002} (\bibinfo {year} {2015})},\ \Eprint
		{http://arxiv.org/abs/1412.1865} {arXiv:1412.1865} \BibitemShut {NoStop}%
		\bibitem [{\citenamefont {Harris}(1996)}]{Harris1996}%
		\BibitemOpen
		\bibfield  {author} {\bibinfo {author} {\bibfnamefont {S.~E.}\ \bibnamefont
				{Harris}},\ }\href {\doibase 10.1103/PhysRevLett.77.5357} {\bibfield
			{journal} {\bibinfo  {journal} {Physical Review Letters}\ }\textbf {\bibinfo
				{volume} {77}},\ \bibinfo {pages} {5357} (\bibinfo {year} {1996})},\ \Eprint
		{http://arxiv.org/abs/1404.0311} {arXiv:1404.0311} \BibitemShut {NoStop}%
		\bibitem [{\citenamefont {Fleischhauer}\ \emph {et~al.}(2005)\citenamefont
			{Fleischhauer}, \citenamefont {Imamoglu},\ and\ \citenamefont
			{Marangos}}]{Fleischhauer2005}%
		\BibitemOpen
		\bibfield  {author} {\bibinfo {author} {\bibfnamefont {M.}~\bibnamefont
				{Fleischhauer}}, \bibinfo {author} {\bibfnamefont {A.}~\bibnamefont
				{Imamoglu}}, \ and\ \bibinfo {author} {\bibfnamefont {J.~P.}\ \bibnamefont
				{Marangos}},\ }\href {\doibase 10.1103/RevModPhys.77.633} {\bibfield
			{journal} {\bibinfo  {journal} {Reviews of Modern Physics}\ }\textbf
			{\bibinfo {volume} {77}},\ \bibinfo {pages} {633} (\bibinfo {year}
			{2005})}\BibitemShut {NoStop}%
		\bibitem [{\citenamefont {Matsko}\ and\ \citenamefont
			{Rostovtsev}(1998)}]{Matsko1998}%
		\BibitemOpen
		\bibfield  {author} {\bibinfo {author} {\bibfnamefont {A.~B.}\ \bibnamefont
				{Matsko}}\ and\ \bibinfo {author} {\bibfnamefont {Y.~V.}\ \bibnamefont
				{Rostovtsev}},\ }\href {\doibase 10.1103/PhysRevE.58.7846} {\bibfield
			{journal} {\bibinfo  {journal} {Physical Review E}\ }\textbf {\bibinfo
				{volume} {58}},\ \bibinfo {pages} {7846} (\bibinfo {year}
			{1998})}\BibitemShut {NoStop}%
		\bibitem [{\citenamefont {Ersfeld}\ and\ \citenamefont
			{Jaroszynski}(2002)}]{Ersfeld2002}%
		\BibitemOpen
		\bibfield  {author} {\bibinfo {author} {\bibfnamefont {B.}~\bibnamefont
				{Ersfeld}}\ and\ \bibinfo {author} {\bibfnamefont {D.~A.}\ \bibnamefont
				{Jaroszynski}},\ }\href {\doibase 10.1080/09500340110111743} {\bibfield
			{journal} {\bibinfo  {journal} {Journal of Modern Optics}\ }\textbf {\bibinfo
				{volume} {49}},\ \bibinfo {pages} {889} (\bibinfo {year} {2002})}\BibitemShut
		{NoStop}%
		\bibitem [{\citenamefont {Hur}\ \emph {et~al.}(2003)\citenamefont {Hur},
			\citenamefont {Wurtele},\ and\ \citenamefont {Shvets}}]{Hur2003}%
		\BibitemOpen
		\bibfield  {author} {\bibinfo {author} {\bibfnamefont {M.~S.}\ \bibnamefont
				{Hur}}, \bibinfo {author} {\bibfnamefont {J.~S.}\ \bibnamefont {Wurtele}}, \
			and\ \bibinfo {author} {\bibfnamefont {G.}~\bibnamefont {Shvets}},\ }\href
		{\doibase 10.1063/1.1580816} {\bibfield  {journal} {\bibinfo  {journal}
				{Physics of Plasmas}\ }\textbf {\bibinfo {volume} {10}},\ \bibinfo {pages}
			{3004} (\bibinfo {year} {2003})}\BibitemShut {NoStop}%
		\bibitem [{\citenamefont {Gordon}\ \emph
			{et~al.}(2000{\natexlab{a}})\citenamefont {Gordon}, \citenamefont {Mori},\
			and\ \citenamefont {Joshi}}]{Gordon2000}%
		\BibitemOpen
		\bibfield  {author} {\bibinfo {author} {\bibfnamefont {D.~F.}\ \bibnamefont
				{Gordon}}, \bibinfo {author} {\bibfnamefont {W.~B.}\ \bibnamefont {Mori}}, \
			and\ \bibinfo {author} {\bibfnamefont {C.}~\bibnamefont {Joshi}},\ }\href
		{\doibase 10.1063/1.874178} {\bibfield  {journal} {\bibinfo  {journal}
				{Physics of Plasmas}\ }\textbf {\bibinfo {volume} {7}},\ \bibinfo {pages}
			{3145} (\bibinfo {year} {2000}{\natexlab{a}})}\BibitemShut {NoStop}%
		\bibitem [{\citenamefont {Gordon}\ \emph
			{et~al.}(2000{\natexlab{b}})\citenamefont {Gordon}, \citenamefont {Mori},\
			and\ \citenamefont {Joshi}}]{Gordon2000a}%
		\BibitemOpen
		\bibfield  {author} {\bibinfo {author} {\bibfnamefont {D.~F.}\ \bibnamefont
				{Gordon}}, \bibinfo {author} {\bibfnamefont {W.~B.}\ \bibnamefont {Mori}}, \
			and\ \bibinfo {author} {\bibfnamefont {C.}~\bibnamefont {Joshi}},\ }\href
		{\doibase 10.1063/1.874179} {\bibfield  {journal} {\bibinfo  {journal}
				{Physics of Plasmas}\ }\textbf {\bibinfo {volume} {7}},\ \bibinfo {pages}
			{3156} (\bibinfo {year} {2000}{\natexlab{b}})}\BibitemShut {NoStop}%
		\bibitem [{\citenamefont {Arefiev}\ \emph {et~al.}(2020)\citenamefont
			{Arefiev}, \citenamefont {Stark}, \citenamefont {Toncian},\ and\
			\citenamefont {Murakami}}]{Arefiev2020}%
		\BibitemOpen
		\bibfield  {author} {\bibinfo {author} {\bibfnamefont {A.}~\bibnamefont
				{Arefiev}}, \bibinfo {author} {\bibfnamefont {D.~J.}\ \bibnamefont {Stark}},
			\bibinfo {author} {\bibfnamefont {T.}~\bibnamefont {Toncian}}, \ and\
			\bibinfo {author} {\bibfnamefont {M.}~\bibnamefont {Murakami}},\ }\href
		{\doibase 10.1063/5.0008018} {\bibfield  {journal} {\bibinfo  {journal}
				{Physics of Plasmas}\ }\textbf {\bibinfo {volume} {27}} (\bibinfo {year}
			{2020}),\ 10.1063/5.0008018},\ \Eprint {http://arxiv.org/abs/2005.12435}
		{arXiv:2005.12435} \BibitemShut {NoStop}%
		\bibitem [{\citenamefont {Tabak}\ \emph {et~al.}(1994)\citenamefont {Tabak},
			\citenamefont {Hammer}, \citenamefont {Glinsky}, \citenamefont {Kruer},
			\citenamefont {Wilks}, \citenamefont {Woodworth}, \citenamefont {Campbell},
			\citenamefont {Perry},\ and\ \citenamefont {Mason}}]{Tabak1994}%
		\BibitemOpen
		\bibfield  {author} {\bibinfo {author} {\bibfnamefont {M.}~\bibnamefont
				{Tabak}}, \bibinfo {author} {\bibfnamefont {J.}~\bibnamefont {Hammer}},
			\bibinfo {author} {\bibfnamefont {M.~E.}\ \bibnamefont {Glinsky}}, \bibinfo
			{author} {\bibfnamefont {W.~L.}\ \bibnamefont {Kruer}}, \bibinfo {author}
			{\bibfnamefont {S.~C.}\ \bibnamefont {Wilks}}, \bibinfo {author}
			{\bibfnamefont {J.}~\bibnamefont {Woodworth}}, \bibinfo {author}
			{\bibfnamefont {E.~M.}\ \bibnamefont {Campbell}}, \bibinfo {author}
			{\bibfnamefont {M.~D.}\ \bibnamefont {Perry}}, \ and\ \bibinfo {author}
			{\bibfnamefont {R.~J.}\ \bibnamefont {Mason}},\ }\href {\doibase
			10.1063/1.870664} {\bibfield  {journal} {\bibinfo  {journal} {Physics of
					Plasmas}\ }\textbf {\bibinfo {volume} {1}},\ \bibinfo {pages} {1626}
			(\bibinfo {year} {1994})}\BibitemShut {NoStop}%
		\bibitem [{\citenamefont {Wang}\ \emph
			{et~al.}(2015{\natexlab{a}})\citenamefont {Wang}, \citenamefont {Gibbon},
			\citenamefont {Sheng},\ and\ \citenamefont {Li}}]{Wang2015b}%
		\BibitemOpen
		\bibfield  {author} {\bibinfo {author} {\bibfnamefont {W.~M.}\ \bibnamefont
				{Wang}}, \bibinfo {author} {\bibfnamefont {P.}~\bibnamefont {Gibbon}},
			\bibinfo {author} {\bibfnamefont {Z.~M.}\ \bibnamefont {Sheng}}, \ and\
			\bibinfo {author} {\bibfnamefont {Y.~T.}\ \bibnamefont {Li}},\ }\href
		{\doibase 10.1103/PhysRevLett.114.015001} {\bibfield  {journal} {\bibinfo
				{journal} {Physical Review Letters}\ }\textbf {\bibinfo {volume} {114}},\
			\bibinfo {pages} {015001} (\bibinfo {year} {2015}{\natexlab{a}})}\BibitemShut
		{NoStop}%
		\bibitem [{\citenamefont {Kitagawa}\ \emph {et~al.}(2022)\citenamefont
			{Kitagawa}, \citenamefont {Mori}, \citenamefont {Ishii}, \citenamefont
			{Hanayama}, \citenamefont {Okihara}, \citenamefont {Arikawa}, \citenamefont
			{Abe}, \citenamefont {Miura}, \citenamefont {Ozaki}, \citenamefont {Komeda},
			\citenamefont {Suto}, \citenamefont {Umetani}, \citenamefont {Sunahra},
			\citenamefont {Johzaki}, \citenamefont {Sakagami}, \citenamefont {Iwamoto},
			\citenamefont {Sentoku}, \citenamefont {Nakajima}, \citenamefont {Sakata},
			\citenamefont {Matsuo}, \citenamefont {Mirfayzi}, \citenamefont {Kawanaka},
			\citenamefont {Fujiokua}, \citenamefont {Tsubakimoto}, \citenamefont
			{Shigemori}, \citenamefont {Yamanoi}, \citenamefont {Yogo}, \citenamefont
			{Nakao}, \citenamefont {Asano}, \citenamefont {Shiraga}, \citenamefont
			{Motohiro}, \citenamefont {Hioki},\ and\ \citenamefont
			{Azuma}}]{Kitagawa2022}%
		\BibitemOpen
		\bibfield  {author} {\bibinfo {author} {\bibfnamefont {Y.}~\bibnamefont
				{Kitagawa}}, \bibinfo {author} {\bibfnamefont {Y.}~\bibnamefont {Mori}},
			\bibinfo {author} {\bibfnamefont {K.}~\bibnamefont {Ishii}}, \bibinfo
			{author} {\bibfnamefont {R.}~\bibnamefont {Hanayama}}, \bibinfo {author}
			{\bibfnamefont {S.}~\bibnamefont {Okihara}}, \bibinfo {author} {\bibfnamefont
				{Y.}~\bibnamefont {Arikawa}}, \bibinfo {author} {\bibfnamefont
				{Y.}~\bibnamefont {Abe}}, \bibinfo {author} {\bibfnamefont {E.}~\bibnamefont
				{Miura}}, \bibinfo {author} {\bibfnamefont {T.}~\bibnamefont {Ozaki}},
			\bibinfo {author} {\bibfnamefont {O.}~\bibnamefont {Komeda}}, \bibinfo
			{author} {\bibfnamefont {H.}~\bibnamefont {Suto}}, \bibinfo {author}
			{\bibfnamefont {Y.}~\bibnamefont {Umetani}}, \bibinfo {author} {\bibfnamefont
				{A.}~\bibnamefont {Sunahra}}, \bibinfo {author} {\bibfnamefont
				{T.}~\bibnamefont {Johzaki}}, \bibinfo {author} {\bibfnamefont
				{H.}~\bibnamefont {Sakagami}}, \bibinfo {author} {\bibfnamefont
				{A.}~\bibnamefont {Iwamoto}}, \bibinfo {author} {\bibfnamefont
				{Y.}~\bibnamefont {Sentoku}}, \bibinfo {author} {\bibfnamefont
				{N.}~\bibnamefont {Nakajima}}, \bibinfo {author} {\bibfnamefont
				{S.}~\bibnamefont {Sakata}}, \bibinfo {author} {\bibfnamefont
				{K.}~\bibnamefont {Matsuo}}, \bibinfo {author} {\bibfnamefont {R.~S.}\
				\bibnamefont {Mirfayzi}}, \bibinfo {author} {\bibfnamefont {J.}~\bibnamefont
				{Kawanaka}}, \bibinfo {author} {\bibfnamefont {S.}~\bibnamefont {Fujiokua}},
			\bibinfo {author} {\bibfnamefont {K.}~\bibnamefont {Tsubakimoto}}, \bibinfo
			{author} {\bibfnamefont {K.}~\bibnamefont {Shigemori}}, \bibinfo {author}
			{\bibfnamefont {K.}~\bibnamefont {Yamanoi}}, \bibinfo {author} {\bibfnamefont
				{A.}~\bibnamefont {Yogo}}, \bibinfo {author} {\bibfnamefont {A.}~\bibnamefont
				{Nakao}}, \bibinfo {author} {\bibfnamefont {M.}~\bibnamefont {Asano}},
			\bibinfo {author} {\bibfnamefont {H.}~\bibnamefont {Shiraga}}, \bibinfo
			{author} {\bibfnamefont {T.}~\bibnamefont {Motohiro}}, \bibinfo {author}
			{\bibfnamefont {T.}~\bibnamefont {Hioki}}, \ and\ \bibinfo {author}
			{\bibfnamefont {H.}~\bibnamefont {Azuma}},\ }\href {\doibase
			10.1088/1741-4326/ac7966} {\bibfield  {journal} {\bibinfo  {journal} {Nuclear
					Fusion}\ }\textbf {\bibinfo {volume} {62}} (\bibinfo {year} {2022}),\
			10.1088/1741-4326/ac7966}\BibitemShut {NoStop}%
		\bibitem [{\citenamefont {Morace}\ \emph {et~al.}(2019)\citenamefont {Morace},
			\citenamefont {Iwata}, \citenamefont {Sentoku}, \citenamefont {Mima},
			\citenamefont {Arikawa}, \citenamefont {Yogo}, \citenamefont {Andreev},
			\citenamefont {Tosaki}, \citenamefont {Vaisseau}, \citenamefont {Abe},
			\citenamefont {Kojima}, \citenamefont {Sakata}, \citenamefont {Hata},
			\citenamefont {Lee}, \citenamefont {Matsuo}, \citenamefont {Kamitsukasa},
			\citenamefont {Norimatsu}, \citenamefont {Kawanaka}, \citenamefont {Tokita},
			\citenamefont {Miyanaga}, \citenamefont {Shiraga}, \citenamefont {Sakawa},
			\citenamefont {Nakai}, \citenamefont {Nishimura}, \citenamefont {Azechi},
			\citenamefont {Fujioka},\ and\ \citenamefont {Kodama}}]{Morace2019}%
		\BibitemOpen
		\bibfield  {author} {\bibinfo {author} {\bibfnamefont {A.}~\bibnamefont
				{Morace}}, \bibinfo {author} {\bibfnamefont {N.}~\bibnamefont {Iwata}},
			\bibinfo {author} {\bibfnamefont {Y.}~\bibnamefont {Sentoku}}, \bibinfo
			{author} {\bibfnamefont {K.}~\bibnamefont {Mima}}, \bibinfo {author}
			{\bibfnamefont {Y.}~\bibnamefont {Arikawa}}, \bibinfo {author} {\bibfnamefont
				{A.}~\bibnamefont {Yogo}}, \bibinfo {author} {\bibfnamefont {A.}~\bibnamefont
				{Andreev}}, \bibinfo {author} {\bibfnamefont {S.}~\bibnamefont {Tosaki}},
			\bibinfo {author} {\bibfnamefont {X.}~\bibnamefont {Vaisseau}}, \bibinfo
			{author} {\bibfnamefont {Y.}~\bibnamefont {Abe}}, \bibinfo {author}
			{\bibfnamefont {S.}~\bibnamefont {Kojima}}, \bibinfo {author} {\bibfnamefont
				{S.}~\bibnamefont {Sakata}}, \bibinfo {author} {\bibfnamefont
				{M.}~\bibnamefont {Hata}}, \bibinfo {author} {\bibfnamefont {S.}~\bibnamefont
				{Lee}}, \bibinfo {author} {\bibfnamefont {K.}~\bibnamefont {Matsuo}},
			\bibinfo {author} {\bibfnamefont {N.}~\bibnamefont {Kamitsukasa}}, \bibinfo
			{author} {\bibfnamefont {T.}~\bibnamefont {Norimatsu}}, \bibinfo {author}
			{\bibfnamefont {J.}~\bibnamefont {Kawanaka}}, \bibinfo {author}
			{\bibfnamefont {S.}~\bibnamefont {Tokita}}, \bibinfo {author} {\bibfnamefont
				{N.}~\bibnamefont {Miyanaga}}, \bibinfo {author} {\bibfnamefont
				{H.}~\bibnamefont {Shiraga}}, \bibinfo {author} {\bibfnamefont
				{Y.}~\bibnamefont {Sakawa}}, \bibinfo {author} {\bibfnamefont
				{M.}~\bibnamefont {Nakai}}, \bibinfo {author} {\bibfnamefont
				{H.}~\bibnamefont {Nishimura}}, \bibinfo {author} {\bibfnamefont
				{H.}~\bibnamefont {Azechi}}, \bibinfo {author} {\bibfnamefont
				{S.}~\bibnamefont {Fujioka}}, \ and\ \bibinfo {author} {\bibfnamefont
				{R.}~\bibnamefont {Kodama}},\ }\href {\doibase 10.1038/s41467-019-10997-1}
		{\bibfield  {journal} {\bibinfo  {journal} {Nature Communications}\ }\textbf
			{\bibinfo {volume} {10}},\ \bibinfo {pages} {1} (\bibinfo {year}
			{2019})}\BibitemShut {NoStop}%
		\bibitem [{\citenamefont {Zhou}\ \emph {et~al.}(2022)\citenamefont {Zhou},
			\citenamefont {Wang}, \citenamefont {Li},\ and\ \citenamefont
			{Zhang}}]{Zhou2022}%
		\BibitemOpen
		\bibfield  {author} {\bibinfo {author} {\bibfnamefont {G.}~\bibnamefont
				{Zhou}}, \bibinfo {author} {\bibfnamefont {W.-m.}\ \bibnamefont {Wang}},
			\bibinfo {author} {\bibfnamefont {Y.}~\bibnamefont {Li}}, \ and\ \bibinfo
			{author} {\bibfnamefont {J.}~\bibnamefont {Zhang}},\ }\href {\doibase
			10.1063/5.0076203} {\bibfield  {journal} {\bibinfo  {journal} {Physics of
					Plasmas}\ }\textbf {\bibinfo {volume} {29}},\ \bibinfo {pages} {052704}
			(\bibinfo {year} {2022})}\BibitemShut {NoStop}%
		\bibitem [{\citenamefont {Scott}\ \emph {et~al.}(2013)\citenamefont {Scott},
			\citenamefont {P{\'{e}}rez}, \citenamefont {Streeter}, \citenamefont {Clark},
			\citenamefont {Davies}, \citenamefont {Schlenvoigt}, \citenamefont {Santos},
			\citenamefont {Hulin}, \citenamefont {Lancaster}, \citenamefont {Dorchies},
			\citenamefont {Fourment}, \citenamefont {Vauzour}, \citenamefont {Soloviev},
			\citenamefont {Baton}, \citenamefont {Rose},\ and\ \citenamefont
			{Norreys}}]{Scott2013}%
		\BibitemOpen
		\bibfield  {author} {\bibinfo {author} {\bibfnamefont {R.~H.}\ \bibnamefont
				{Scott}}, \bibinfo {author} {\bibfnamefont {F.}~\bibnamefont {P{\'{e}}rez}},
			\bibinfo {author} {\bibfnamefont {M.~J.}\ \bibnamefont {Streeter}}, \bibinfo
			{author} {\bibfnamefont {E.~L.}\ \bibnamefont {Clark}}, \bibinfo {author}
			{\bibfnamefont {J.~R.}\ \bibnamefont {Davies}}, \bibinfo {author}
			{\bibfnamefont {H.~P.}\ \bibnamefont {Schlenvoigt}}, \bibinfo {author}
			{\bibfnamefont {J.~J.}\ \bibnamefont {Santos}}, \bibinfo {author}
			{\bibfnamefont {S.}~\bibnamefont {Hulin}}, \bibinfo {author} {\bibfnamefont
				{K.~L.}\ \bibnamefont {Lancaster}}, \bibinfo {author} {\bibfnamefont
				{F.}~\bibnamefont {Dorchies}}, \bibinfo {author} {\bibfnamefont
				{C.}~\bibnamefont {Fourment}}, \bibinfo {author} {\bibfnamefont
				{B.}~\bibnamefont {Vauzour}}, \bibinfo {author} {\bibfnamefont {A.~A.}\
				\bibnamefont {Soloviev}}, \bibinfo {author} {\bibfnamefont {S.~D.}\
				\bibnamefont {Baton}}, \bibinfo {author} {\bibfnamefont {S.~J.}\ \bibnamefont
				{Rose}}, \ and\ \bibinfo {author} {\bibfnamefont {P.~A.}\ \bibnamefont
				{Norreys}},\ }\href {\doibase 10.1088/1367-2630/15/9/093021} {\bibfield
			{journal} {\bibinfo  {journal} {New Journal of Physics}\ }\textbf {\bibinfo
				{volume} {15}} (\bibinfo {year} {2013}),\
			10.1088/1367-2630/15/9/093021}\BibitemShut {NoStop}%
		\bibitem [{\citenamefont {Zhang}\ \emph {et~al.}(2023)\citenamefont {Zhang},
			\citenamefont {Wang}, \citenamefont {Li},\ and\ \citenamefont
			{Zhang}}]{Zhang2023}%
		\BibitemOpen
		\bibfield  {author} {\bibinfo {author} {\bibfnamefont {T.-H.}\ \bibnamefont
				{Zhang}}, \bibinfo {author} {\bibfnamefont {W.-M.}\ \bibnamefont {Wang}},
			\bibinfo {author} {\bibfnamefont {Y.-T.}\ \bibnamefont {Li}}, \ and\ \bibinfo
			{author} {\bibfnamefont {J.}~\bibnamefont {Zhang}},\ }\href
		{http://arxiv.org/abs/2311.11625} {\  (\bibinfo {year} {2023})},\ \Eprint
		{http://arxiv.org/abs/2311.11625} {arXiv:2311.11625} \BibitemShut {NoStop}%
		\bibitem [{\citenamefont {Wang}\ \emph
			{et~al.}(2015{\natexlab{b}})\citenamefont {Wang}, \citenamefont {Gibbon},
			\citenamefont {Sheng},\ and\ \citenamefont {Li}}]{Wang2015a}%
		\BibitemOpen
		\bibfield  {author} {\bibinfo {author} {\bibfnamefont {W.~M.}\ \bibnamefont
				{Wang}}, \bibinfo {author} {\bibfnamefont {P.}~\bibnamefont {Gibbon}},
			\bibinfo {author} {\bibfnamefont {Z.~M.}\ \bibnamefont {Sheng}}, \ and\
			\bibinfo {author} {\bibfnamefont {Y.~T.}\ \bibnamefont {Li}},\ }\href
		{\doibase 10.1103/PhysRevE.91.013101} {\bibfield  {journal} {\bibinfo
				{journal} {Physical Review E}\ }\textbf {\bibinfo {volume} {91}},\ \bibinfo {pages} {013101} (\bibinfo
			{year} {2015}{\natexlab{b}})},\ \Eprint {http://arxiv.org/abs/1409.1808}
		{arXiv:1409.1808} \BibitemShut {NoStop}%
		\bibitem [{Sup()}]{SupMat}%
		\BibitemOpen
		\href@noop {} {}\bibinfo {note} {See Supplemental Material, which includes Refs. [36, 37]}\BibitemShut
		{NoStop}%
		\bibitem [{\citenamefont {Cattani}\ \emph {et~al.}(2000)\citenamefont
			{Cattani}, \citenamefont {Kim}, \citenamefont {Anderson},\ and\ \citenamefont
			{Lisak}}]{Cattani2000}%
		\BibitemOpen
		\bibfield  {author} {\bibinfo {author} {\bibfnamefont {F.}~\bibnamefont
				{Cattani}}, \bibinfo {author} {\bibfnamefont {A.}~\bibnamefont {Kim}},
			\bibinfo {author} {\bibfnamefont {D.}~\bibnamefont {Anderson}}, \ and\
			\bibinfo {author} {\bibfnamefont {M.}~\bibnamefont {Lisak}},\ }\href
		{\doibase 10.1103/PhysRevE.62.1234} {\bibfield  {journal} {\bibinfo
				{journal} {Physical Review E}\ }\textbf {\bibinfo {volume} {62}},\ \bibinfo
			{pages} {1234} (\bibinfo {year} {2000})}\BibitemShut {NoStop}%
		\bibitem [{\citenamefont {Siminos}\ \emph {et~al.}(2012)\citenamefont
			{Siminos}, \citenamefont {Grech}, \citenamefont {Skupin}, \citenamefont
			{Schlegel},\ and\ \citenamefont {Tikhonchuk}}]{Siminos2012}%
		\BibitemOpen
		\bibfield  {author} {\bibinfo {author} {\bibfnamefont {E.}~\bibnamefont
				{Siminos}}, \bibinfo {author} {\bibfnamefont {M.}~\bibnamefont {Grech}},
			\bibinfo {author} {\bibfnamefont {S.}~\bibnamefont {Skupin}}, \bibinfo
			{author} {\bibfnamefont {T.}~\bibnamefont {Schlegel}}, \ and\ \bibinfo
			{author} {\bibfnamefont {V.~T.}\ \bibnamefont {Tikhonchuk}},\ }\href
		{\doibase 10.1103/PhysRevE.86.056404} {\bibfield  {journal} {\bibinfo
				{journal} {Physical Review E}\ }\textbf {\bibinfo {volume} {86}},\ \bibinfo
			{pages} {056404} (\bibinfo {year} {2012})},\ \Eprint
		{http://arxiv.org/abs/1209.3322} {arXiv:1209.3322} \BibitemShut {NoStop}%
	\end{thebibliography}
	\bibliographystyle{apsrev4-1} 
\end{document}